\def\limepy{\textsc{limepy}}

\def\python{\textsc{python}}
\def\emcee{\textsc{emcee}}

\def\intfrac[#1,#2]{I_{#1}^{#2}}
\def\derfrac[#1,#2]{D_{#1}^{#2}}

\def\Eg{E_\gamma}

\def\rh{r_{\rm h}}

\def\ra{r_{\rm a}}

\def\rcore{R_{\rm c}}

\def\rt{r_{\rm t}}

\def\phit{\phit_{\rm t}}

\def\msun{\mbox{M}_{\rm \odot}}
\def\trh{\tau_{\rm rh}}
\def\teq{\tau_{\rm eq}}

\documentclass[fleqn,usenatbib]{mnras}
\usepackage{amssymb,latexsym,graphicx,natbib,eufrak,times,amsmath,xfrac,nicefrac}
\usepackage[caption=false]{subfig}
\usepackage{wrapfig}

\title[The effect of stellar-mass black holes on the central kinematics of $\omega$ Cen]{The effect of stellar-mass black holes on the central kinematics of $\omega$ Cen: a cautionary tale for IMBH interpretations}
\author[Alice Zocchi, Mark Gieles, Vincent H\'enault-Brunet]
  {Alice Zocchi$^{1,2}$\thanks{E-mail:
azocchi@cosmos.esa.int}, Mark Gieles$^{3}$, Vincent H\'enault-Brunet$^{4,5}$ \\
$^1$ European Space Research and Technology Centre (ESA/ESTEC), Keplerlaan 1, 2201 AZ Noordwijk, The Netherlands. \\
$^2$ Dipartimento di Fisica e Astronomia, Universit\`{a} degli Studi di Bologna, via Gobetti 93/2, I–40127, Bologna, Italy. \\
$^3$ Department of Physics, University of Surrey, Guildford, GU2 7XH,UK. \\
$^4$ National Research Council, Herzberg Astronomy \& Astrophysics, 5071 West Saanich Road, Victoria, BC, V9E 2E7, Canada. \\
$^5$ Department of Astrophysics/IMAPP, Radboud University, PO Box 9010, NL-6500 GL Nijmegen, the Netherlands. \\
}
\date{Accepted 2018 May 26. Received 2018 April 25; in original form 2017 September 15.}

\def\LaTeX{L\kern-.36em\raise.3ex\hbox{a}\kern-.15em
    T\kern-.1667em\lower.7ex\hbox{E}\kern-.125emX}

\begin{document}         

\label{firstpage}
\pagerange{\pageref{firstpage}--\pageref{lastpage}}
\maketitle

\begin{abstract}
The search for intermediate-mass black holes (IMBHs) in the centre of globular clusters is often based on the observation of a central cusp in the surface brightness profile and a rise towards the centre in the velocity dispersion profiles. Similar signatures, however, could result from other effects, that need to be taken into account in order to determine the presence (or the absence) of an IMBH in these stellar systems. Following our previous exploration of the role of radial anisotropy in shaping these observational signatures, we analyse here the effects produced by the presence of a population of centrally concentrated stellar-mass black holes. We fit dynamical models to $\omega$ Cen data, and we show that models with $\sim 5\%$ of their mass in black holes (consistent with $\sim 100\%$ retention fraction after natal kicks) can reproduce the data. When simultaneously considering both radial anisotropy and mass segregation, the best-fit model includes a smaller population of remnants, and a less extreme degree of anisotropy with respect to the models that include only one of these features. These results underline that before conclusions about putative IMBHs can be made, the effects of stellar-mass black holes and radial anisotropy need to be properly accounted for.
\end{abstract}

\begin{keywords}
methods: numerical, stars: kinematics and dynamics, globular clusters: general, globular clusters: individual: $\omega$ Centauri (NGC 5139), galaxies: star clusters: general
\end{keywords}


\section{Introduction}

Intermediate-mass black holes (IMBHs) could provide the missing link to understand the origin of supermassive black holes and of their host galaxies \citep{Ebisuzaki2001}. It has been suggested that IMBHs could form via a runaway stellar collision process in young ($\lesssim 2$~Myr) massive star clusters \citep{PortegiesZwart2004,Gieles2018SMS}, and great observational effort has been devoted to finding them in dense stellar systems like globular clusters \citep[see, e.g.,][]{LuKong2011,Nora2011,Strader2012,Haggard2013,Luetz2013,Feldmeier2013}. The difficulty is that several other factors could cause observational signatures compatible with the presence of an IMBH \citep{Vesperini2010,vdMA2010}. It is therefore important to determine the effect of these alternative ingredients on the quantities that are usually observed for these systems, in order to establish if an IMBH is indeed present or not.

The most common observational signature used to infer the presence of an IMBH in globular clusters is a rise towards the centre in the velocity dispersion profiles. \citet{Baumgardt2004} carried out numerical simulations of star clusters with a central IMBH and with a realistic mass function, taking also into account stellar evolution. They showed that the presence of the IMBH produces a cusp in the velocity dispersion profile in the innermost region of the cluster (within $\sim 0.01$ half-mass radii, $r_{\rm h}$, in their models), and causes the velocity dispersion profile to be larger than what can be explained by the stars over a radius of about $\sim 0.1 r_{\rm h}$. This is also observed by \citet{Baumgardt2017} in the case of the best-fit model obtained for the globular cluster $\omega$ Cen (see their Fig.~6). In a recent work \citep{Zocchi2017}, we showed that radially anisotropic models reproduce the observational profiles of $\omega$ Cen well, and describe the central kinematics as derived from Hubble Space Telescope (\textit{HST}) proper motions without the need for an IMBH. In this paper we consider another factor that could affect the central velocity dispersion in a similar way as an IMBH, namely a centrally concentrated population of stellar-mass black holes (BHs). We point out that both radial anisotropy and the presence of a population of BHs produce an increase in the central projected velocity dispersion without generating a cusp like the one expected from the presence of an IMBH, but the data available at the moment does not allow to discriminate between these effects.

The possibility that old globular clusters host stellar-mass BHs has historically received little attention. This is, firstly, because BHs are believed to experience a natal kick at their formation in the supernova explosion, which could bring them to a velocity larger than the escape velocity. However, little is known about the magnitude of the corresponding kick velocity, because constraining this empirically has proven challenging. Recent efforts have taken advantage of the distribution in the Milky Way of X-ray binaries that contain black holes \citep[BH-XRBs;][]{Repettoetal2012,Repetto2015,Mandel2016,Repetto2017}: the analyses based on the sample of observed BH-XRBs reveal that some systems could be explained with no or small natal kicks, but some others are better described when considering a relatively large natal kick. However, these analyses admittedly do not account for the fact that the presence of BH-XRBs found at higher Galactic latitude could be explained by considering the possibility that a few systems have been formed in the halo, or that they could have been ejected from globular clusters by dynamical interactions. 

Secondly, the (unknown) fraction of BHs that is retained after supernova kicks was believed to be ejected quickly due to dynamical interactions. \citet{Spitzer1969} showed that for a stellar system composed of two stellar populations with masses $m_1$ and $m_2$, where $m_1 < m_2$, and $M_1$ and $M_2$ the total mass of the two populations, with $M_1 \gg M_2$, equipartition is only possible if:
\begin{equation}
M_2 < 0.16 M_1 \left(\frac{m_1}{m_2}\right)^{3/2} 
\label{Spitzer}
\end{equation}
\citep[see also][]{Watters2000}.
If this condition\footnote{This condition was later generalised for a continuous mass spectrum by \citet{Vishniac1978}.} is not satisfied, heavy objects (e.g., BHs) become self-gravitating before equipartition is achieved: they form a compact subsystem in the centre, dynamically separated from the rest of the cluster, and they interact only with each other. Due to the short two-body relaxation timescale of such a subsystem, dynamical ejections are very efficient. Therefore, it was often assumed that BHs quickly eject each other from the cluster, until a single BH-binary is left.

The interest in the dynamical behaviour of a BH subsystems in globular clusters was recently reignited by the discovery of BH candidates in several globular clusters with radio \citep{Strader2012b,Chomiuketal2013,MillerJonesetal2015} and X-ray observations \citep{Maccarone2007}. \citet{BreenHeggie2013} showed that the dynamical ejection rate of BHs is lower than what was generally assumed \citep[see also][]{Morscher2013,Morscher2015}. The $N$-body models presented by \citet{BreenHeggie2013} demonstrate that a BH subsystem can survive as long as $\sim 10 \trh$, where $\trh$ is the half-mass radius relaxation time. From a comparison of multi-component dynamical models to $N$-body models, \citet{Peuten2017} showed that the BHs do not achieve equipartition with the stars. This means that if natal kicks are low enough for BHs to be retained by the cluster, and the initial half-mass radius relaxation time is long enough, a BH population is expected to be present in stellar systems as old as globular clusters. In fact, \citet{BreenHeggie2013} suggest that the collapse of the visible core coincides with the moment when all BHs have escaped the system. Given that only $\sim20\%$ of the Milky Way globular clusters are classified as core collapsed \citep{DjorgovskiKing1986}, this may mean that a fraction as high as 80\% of Galactic globular clusters still contain a population of BHs.

With this in mind, it is worth considering what the effects are of a BH population on the rest of the stars. \citet{Merrittetal2004} showed that a population of heavy dark remnants inflates the core radius ($\rcore$) measured from the visible stars in the cluster. \citet{Mackeyetal2008} suggested that the observed increase of $\rcore$ with age in clusters in the Large Magellanic Cloud can be explained by a large retention fraction of BHs (i.e. low kick velocities). Finally, \citet{Peuten2016} showed that the distribution of stars of different masses in the Galactic globular cluster NGC\,6101, which displays a surprising lack of mass segregation \citep{D15}, can be reproduced by $N$-body models and dynamical multimass models in which $\gtrsim 50\%$ of the BHs are retained after supernova explosions \citep[see also][]{Alessandrini2016}.

$\omega$ Cen is a likely candidate to host a BH population at the present day. Because of its large mass, it had a large escape velocity at the time BHs formed. Its present half-mass radius relaxation time is $\trh \sim20\,$Gyr, longer than its age. The relaxation time must have been shorter in the past, because of some expansion following stellar mass loss, so the dynamical age of $\omega$ Cen just falls short of one $\trh$. During this time, we expect $\lesssim10\%$ of the BHs retained after supernova kicks to be ejected dynamically \citep{BreenHeggie2013}. For the metallicity of $\omega$ Cen, the stellar evolution models of \citet{Hurley2000}, SSE, predict that about 5\% of the present day mass is in the form of stellar-mass BHs for a \citet{Kroupa2001} stellar initial mass function (IMF) between $0.1\,\msun$ and $100\,\msun$. If a third of those are lost in supernova kicks and dynamical ejections \citep{Morscher2015}, we estimate that $\omega$ Cen hosts $\sim2\times10^4$ stellar-mass BHs (for BHs with a mass of $5\,\msun$ and a mass for $\omega$ Cen of $\sim3\times10^6\,\msun$, as estimated in \citealt{Zocchi2017}). \citet{Spera2015} combined recent stellar evolutionary tracks with models for supernova explosion, and predicted the mass distribution of compact remnants. They compared their results with those obtained by other codes, and found that, interestingly, the fraction of BHs produced from an IMF is remarkably similar across all the codes they compare and almost independent of metallicity. However, there are significant differences when focusing on the massive BHs: at low metallicity ($Z\lesssim0.002$, applicable to most globular clusters), they find significantly larger maximum BH masses compared to previous predictions (e.g. from SSE), and 5-6 times more massive ($>25$ $\msun$) black holes compared to SSE. The typical mass of BHs in $\omega$ Cen may therefore be larger than 5 M$_{\sun}$. Moreover, \citet{ShanahanGieles2015} showed that the total mass in BHs with respect to the total mass of the cluster also depends on metallicity: they found that for $Z = Z_{\sun}$ the BHs account for 4\% of the total mass, while for $Z = 0.01 Z_{\sun}$ they make up 7\% of the total mass.

Despite the young dynamical age of $\omega$ Cen, the BHs should have already segregated to the centre of the cluster. The equipartition timescale of a two-component system depends on $\trh$ as $\teq\simeq(m_1/m_2)\trh$ \citep{Spitzer1969}. Because stellar-mass BHs are an order of magnitude more massive than the stars, it only takes $0.1\trh$ for the BHs to reach the centre, i.e. enough time given the dynamical age of $\sim1\,\trh$ estimated for $\omega$ Cen. A mild mass segregation has been observed among the visible stars of this cluster \citep{Anderson2002,Bellini2018} and by means of an analysis of blue straggler stars \citep{Ferraro2006}, consistent with its current state of partial relaxation \citep{GierszHeggie2003}.

The work we are presenting provides information on the presence of stellar mass BHs in $\omega$ Cen and on their masses from a dynamical point of view. We use here dynamical models with two mass components, to account for the different dynamics of stars and BHs in the presence of mass segregation. This allows us to estimate the amount of mass contained in the invisible BH component, and to determine the effect this has on the observed signatures that are often related to the presence of an IMBH in its centre. The fact that the visible stars are only mildly segregated justifies the choice of considering them all as part of the same component. In addition, by means of numerical simulations, \citet{Peuten2017} showed that in dynamically evolved systems with BHs, mass segregation among the stars is strongly suppressed, and the dynamics of stars, white dwarfs, and neutron stars are very similar to each other, further justifying the approximation by a single component for the stars and low-mass remnants, with visible stars used as ``tracers'' of a larger population.

This paper is organised as follows: in Section~\ref{Sect_Models} we present the dynamical models we use, and in Section~\ref{Sect_Fit} we describe the data and the fitting procedure we adopt. We present and discuss our results in Section~\ref{Sect_Res}, and we draw our conclusions in Section~\ref{Sect_Concl}.


\section{Models}
\label{Sect_Models}

To determine the effect of the presence of a population of dark remnants on the observable quantities used to study the dynamics of globular clusters, we use the \limepy\ family of models\footnote{The \limepy\ (Lowered Isothermal Model Explorer in PYthon) code is available from \href{https://github.com/mgieles/limepy}{https://github.com/mgieles/limepy}.} introduced by \citet{GielesZocchi}, by considering its formulation with multiple mass components. For these models, it is possible to compute several quantities, of the cluster as a whole, and for each component separately.

These models can describe stellar systems with isotropic or anisotropic velocity distributions. The anisotropic version of these models describes systems that are isotropic in the centre, radially anisotropic in their intermediate part, and isotropic again at the truncation. Multiple components can be introduced, each one tracing a population of stars with a given mass, making up a certain fraction of the total mass of the system. The dynamics of each component is determined by the mean mass of the objects it represents, and these models can describe systems with different degrees of mass segregation.

In this study we approximate $\omega$ Cen by two component models: a low-mass component (in the following also called ``visible component'') representing the stars and lower-mass stellar remnants (white dwarfs and neutron stars) and a high-mass component (in the following also called ``dark component'') representing the BHs. These models require only two additional parameters with respect to single component models, and allow us to isolate the role of BHs in determining the dynamics of the other stars.

\subsection{Distribution function}

The \limepy\ family of models is a generalisation of the formulation proposed by \citet{G-L_V2014}, with multiple mass components included as suggested by \citet{DaCostaFreeman1976} and \citet{GunnGriffin1979}. When considering $N_{\rm comp}$ mass components, the models are defined by the sum of $N_{\rm comp}$ distribution functions, each describing a different mass component, and defined as a function of the specific energy $E$ and angular momentum $J$ \citep{GielesZocchi}:
\begin{equation}
f_j(E,J^2) = A_j \, \exp \left(-\frac{J^2}{2 r_{{\rm a},j}^2 s_j^2}\right) \Eg \left(g, \frac{\phi(\rt)-E}{s_j^2}\right)
\label{Eq_DF_Limepy}
\end{equation}
for $E<\phi(\rt)$ and zero otherwise, where $\phi$ is the specific potential, and $\rt$ is the truncation radius. The parameters $s_j$ and $r_{{\rm a},j}$ indicate the velocity scale and the anisotropy radius for the $j$-th component; $A_j$ is a normalization constant. The exponential function introduced in equation~(\ref{Eq_DF_Limepy}) is defined as
\begin{align}
\Eg(a, x) =  
\begin{cases}
\exp(x)  &  a=0 \\
\displaystyle\exp(x) \frac{\gamma(a, x)}{\Gamma(a)}   &  a>0 \ ,
\end{cases}
\label{eq:eg}
\end{align}
where $\gamma(a, x)$ is the lower incomplete gamma function, and $\Gamma(a)$ is the gamma function. 

The models are solved by computing the potential from the Poisson equation:
\begin{equation}
\nabla^2 \phi = 4 \pi G \sum_j \rho_j \ ,
\end{equation}
where 
\begin{equation}
\rho_j = \int f_j  {\rm d}^3v
\end{equation}
is the mass density of the $j$-th component.

\subsection{Model parameters}
\label{Sect_Models_Params}

A model in the family is identified by specifying the values of three parameters. The \textit{concentration parameter} $W_0$ represents the central dimensionless potential; it is a boundary condition for solving the Poisson equation, and determines the shape of the radial profiles of some relevant quantities. The \textit{truncation parameter} $g$ imposes
the sharpness of the truncation in energy \citep[][]{G-L_V2014}: models with large $g$ are more extended, and their truncation is less abrupt.\footnote{To further clarify the role of this parameter, we point out that for isotropic models, by setting $g = 0, 1$ and 2, we respectively obtain the \citet{Woolley1954}, \citet{King1966} and (isotropic, non-rotating) \citet{Wilson1975} models, which have all been used to describe globular clusters.}  The \textit{anisotropy radius} $\ra$ sets the amount of anisotropy in the system, by including it in the same way as in \citet{Michie1963} models: the larger it is, the less radially anisotropic is the model, and when $\ra$ is large with respect to the truncation radius $\rt$, the model is everywhere isotropic. In this paper we will often refer to the dimensionless anisotropy radius $\hat{r}_{\rm a}$, defined as the ratio of the anisotropy radius to the scale radius.

The models are solved in dimensionless units \citep[for a definition, see Section 2.1.2 of][]{GielesZocchi}. To describe every property of the model in the desired set of units, it is necessary to define the velocity, radial and mass scales. These scales are related through the gravitational constant $G$, so that it is enough to specify two scales to completely determine the set of units to use. We consider a mass scale and a radial scale, as these are intuitive quantities to use when fitting the models to data. In particular these scales are adopted to define the total mass of the cluster, $M$, and the half-mass radius, $r_{\rm h}$.

To include multiple mass components, several additional parameters are needed. Each component is defined by setting the value of the mean mass of the stars it describes, $m_j$, and the value of their total mass, $M_j$. In the present paper we consider only two components, one accounting for visible stars and low-mass remnants ($j = 1$), and the other for a population of black holes ($j = 2$); these components are set up by specifying the values of two dimensionless parameters. The first parameter represents the ratio between the mean mass of the dark and visible component, $f_{2,1} = m_2/m_1$, and it is used to determine the mean mass of the dark component as $m_2 = m_1 \times f_{2,1}$. In order to have an estimate of the mean mass of each component expressed in solar masses, we assume $m_1 = 0.3$ M$_{\sun}$. However, the value assumed for $m_1$ has no role in the calculation of the model, and can be easily changed to any other value without affecting the fitting results. The total mass of each component is expressed with respect to the total mass of the cluster: $F_j = M_j/M$. Because $F_1 + F_2 = 1$ and $M$ is defined as one of the two scale parameters (see above), $F_2$ is the second parameter needed to set up the mass function of the models.

The models have two additional parameters, $\delta$ and $\eta$, which set the mass dependence of the velocity scale $s_j$ and of the anisotropy radius of each component $r_{{\rm a},j}$ (in equation~\ref{Eq_DF_Limepy}) with respect to their global values $s$ and $\ra$:
\begin{align}
s_j &= s \left( \frac{m_j}{\bar{m}} \right)^{-\delta} \\
r_{{\rm a},j} &= \ra \left( \frac{m_j}{\bar{m}} \right)^{\eta}
\end{align}
where $\bar{m}$ is the mean mass of the stars in the system. We note that our choice for $\bar{m}$ is different from the one proposed by \citet{DaCostaFreeman1976}, \citet{GunnGriffin1979}, and \citet{GielesZocchi}, where $\bar{m}$ equals the central density weighted mean mass. \citet{Peuten2017} used the same definition used here, with the mean mass calculated simply as the mean mass of all the stars in the model, without any additional weight. In the following we indicate with $W_0$ the concentration parameter computed by using a global mean mass for the stars, while the alternative parameter $W_0^*$ is obtained when considering a central density weighted mass for the objects in the cluster. The \textit{mass segregation parameter} $\delta$ sets the dependence of the velocity scale $s_j$ of each component on the mean mass of its stars. A positive value of this parameter means that more massive stars have smaller velocity scales. Based on the comparison of the \limepy\ models with numerical simulations of clusters with black holes presented by \citet{Peuten2017}, we set the value of this parameter to $\delta = 0.35$; in particular, we note that this value is appropriate for clusters that still contain a large population of BHs. The \textit{anisotropy parameter} $\eta$ allows to modify the anisotropy for the different components. Here we do not explore the role of this parameter, and we simply fix its value to $\eta = 0$, consistent with what found in the numerical simulations of \citet{Peuten2017} for relatively unevolved systems, which correspond roughly to the dynamical age we estimated for $\omega$ Cen. We notice that the choice of $\eta = 0$ does not imply that the different mass components have the same anisotropy.

We notice that the \citet{Spitzer1969} criterion (equation~\ref{Spitzer}) can be expressed with our parameters as:
\begin{equation}
\frac{F_2}{1 - F_2} < 0.16 f_{2,1}^{-3/2} \ .
\end{equation}
When this criterion is not satisfied, equipartition in the cluster is impossible to achieve \citep{Spera2016,Bianchini2016}. Almost all the models considered here violate this criterion, justifying our choice $\delta < 0.5$ \citep[see also][]{Peuten2017}.


\begin{table*}
\begin{center}
\caption[Best-fit parameters for $\omega$ Cen.]{Best-fit parameters for $\omega$ Cen, obtained with the two-steps fitting procedure. Each line refers to a different model, identified by the ID listed in the first column. For each model, we provide the values of $f_{2,1}$ and $F_2$ and the values of the fitting parameters, namely the concentration parameter, with the global ($W_0$) and central ($W_0^*$) definitions described in Section~\ref{Sect_Res_onestep} \citep[see also][]{Peuten2017}, the truncation parameter $g$, the mass $M$ in units of $10^6$ M$_{\sun}$, the half-mass radius $\rh$ in units of pc, and the mass-to-light ratio $M/L$ in solar units. The uncertainties are indicated for the best-fit parameters of each model and for the alternative definition of the concentration parameter, $W_0^*$. In the last column, we list the values of the reduced chi-squared, $\widetilde{\chi}^2 = \chi_{\rm 1}^2/N_{\rm 1} + \chi_{\rm 2}^2/N_{\rm 2}$, calculated as the sum of the reduced chi-squared of the first and second step of the fitting procedure (equations \ref{eq_chi1} and \ref{eq_chi2}).} 
\label{Tab1}
\renewcommand{\arraystretch}{1.25}%
\begin{tabular}{cccccccccc}
\hline\hline
ID & $f_{2,1}$ & $F_2$ & $W_0$ & $W_0^*$ & $g$ & $M$ & $\rh$ & $M/L$ & $\widetilde{\chi}^2$\\
\hline
3\_001  &  3 & 0.001 & $4.78^{+0.95}_{-1.55}$ &   $4.82^{+0.99}_{-1.67}$ & $1.97^{+0.65}_{-0.58}$ & $2.73^{+0.32}_{-0.30}$ & $8.73^{+0.53}_{-0.51}$ & $2.39 \pm 0.03$ & 7.24\\
3\_01   &  3 & 0.01  & $4.62^{+1.06}_{-1.52}$ &   $4.98^{+1.14}_{-1.82}$ & $1.98^{+0.66}_{-0.53}$ & $2.75^{+0.31}_{-0.30}$ & $8.69^{+0.52}_{-0.49}$ & $2.38 \pm 0.03$ & 6.67\\
3\_03   &  3 & 0.03  & $4.54^{+0.91}_{-1.96}$ &   $5.14^{+1.45}_{-2.16}$ & $2.02^{+0.66}_{-0.53}$ & $2.77^{+0.32}_{-0.30}$ & $8.63^{+0.51}_{-0.49}$ & $2.34 \pm 0.03$ & 7.11\\
3\_05   &  3 & 0.05  & $4.35^{+0.93}_{-1.18}$ &   $5.23^{+1.51}_{-2.12}$ & $2.04^{+0.63}_{-0.47}$ & $2.81^{+0.32}_{-0.31}$ & $8.56^{+0.47}_{-0.46}$ & $2.31 \pm 0.03$ & 6.95\\
3\_1    &  3 & 0.1   & $4.04^{+0.71}_{-1.15}$ &   $5.27^{+1.57}_{-2.28}$ & $2.11^{+0.62}_{-0.44}$ & $2.90^{+0.34}_{-0.32}$ & $8.44^{+0.41}_{-0.43}$ & $2.24 \pm 0.02$ & 6.65\\
\hline
5\_001  &  5 & 0.001 & $4.66^{+1.17}_{-1.03}$ &   $4.98^{+1.06}_{-1.67}$ & $1.95^{+0.61}_{-0.50}$ & $2.74 \pm 0.30$        & $8.75^{+0.55}_{-0.51}$ & $2.39 \pm 0.03$ & 6.85\\
5\_01   &  5 & 0.01  & $4.90^{+0.64}_{-1.20}$ &   $6.02^{+1.80}_{-2.47}$ & $1.95^{+0.62}_{-0.57}$ & $2.73^{+0.32}_{-0.30}$ & $8.74^{+0.54}_{-0.51}$ & $2.38 \pm 0.03$ & 5.53\\
5\_03   &  5 & 0.03  & $4.44^{+0.66}_{-1.15}$ &   $6.75^{+2.33}_{-2.96}$ & $2.02^{+0.55}_{-0.39}$ & $2.77^{+0.30}_{-0.28}$ & $8.65^{+0.51}_{-0.49}$ & $2.34 \pm 0.03$ & 5.40\\
5\_05   &  5 & 0.05  & $4.04^{+0.61}_{-0.82}$ &   $6.76^{+2.31}_{-2.96}$ & $2.10^{+0.49}_{-0.35}$ & $2.81^{+0.30}_{-0.29}$ & $8.56 \pm 0.46$        & $2.30 \pm 0.03$ & 5.16\\
5\_1    &  5 & 0.1   & $3.45^{+0.53}_{-1.32}$ &   $6.16^{+2.06}_{-2.73}$ & $2.20^{+0.48}_{-0.33}$ & $2.90^{+0.32}_{-0.31}$ & $8.33^{+0.41}_{-0.43}$ & $2.23 \pm 0.02$ & 5.81\\
\hline
10\_001 & 10 & 0.001 & $4.86^{+0.99}_{-2.08}$ &   $5.79^{+1.53}_{-2.17}$ & $1.91^{+0.63}_{-0.55}$ & $2.73^{+0.32}_{-0.30}$ & $8.78^{+0.56}_{-0.52}$ & $2.40 \pm 0.03$ & 6.50\\
10\_01  & 10 & 0.01  & $4.57^{+0.75}_{-1.40}$ &  $10.54^{+4.12}_{-4.69}$ & $1.97^{+0.48}_{-0.42}$ & $2.75^{+0.28}_{-0.27}$ & $8.81^{+0.52}_{-0.51}$ & $2.37 \pm 0.03$ & 4.85\\
10\_03  & 10 & 0.03  & $3.43^{+0.70}_{-1.14}$ &  $10.58^{+4.14}_{-4.91}$ & $2.22^{+0.35}_{-0.27}$ & $2.80^{+0.28}_{-0.27}$ & $8.68^{+0.47}_{-0.49}$ & $2.32 \pm 0.03$ & 4.00\\
10\_05  & 10 & 0.05  & $2.96^{+0.49}_{-0.75}$ &   $8.84^{+3.41}_{-4.19}$ & $2.32^{+0.35}_{-0.25}$ & $2.83 \pm 0.28$        & $8.52 \pm 0.45$        & $2.28 \pm 0.03$ & 4.35\\
10\_1   & 10 & 0.1   & $2.34^{+0.37}_{-0.53}$ &   $7.16^{+2.59}_{-3.28}$ & $2.36^{+0.35}_{-0.27}$ & $2.92 \pm 0.31$        & $8.28^{+0.40}_{-0.41}$ & $2.21 \pm 0.02$ & 4.96\\
\hline
20\_001 & 20 & 0.001 & $5.08^{+0.84}_{-1.71}$ &   $9.00^{+3.17}_{-4.07}$ & $1.88^{+0.62}_{-0.51}$ & $2.72^{+0.31}_{-0.29}$ & $8.78^{+0.53}_{-0.50}$ & $2.40 \pm 0.03$ & 5.59\\
20\_01  & 20 & 0.01  & $3.97^{+0.63}_{-1.19}$ &  $19.70^{+6.81}_{-8.50}$ & $2.20^{+0.32}_{-0.26}$ & $2.79^{+0.28}_{-0.27}$ & $8.73^{+0.41}_{-0.44}$ & $2.35 \pm 0.03$ & 3.71\\
20\_03  & 20 & 0.03  & $2.55^{+0.48}_{-0.74}$ &  $14.44^{+5.08}_{-7.80}$ & $2.45^{+0.27}_{-0.19}$ & $2.84 \pm 0.27$        & $8.67^{+0.43}_{-0.46}$ & $2.29 \pm 0.03$ & 3.50\\
20\_05  & 20 & 0.05  & $1.96^{+0.35}_{-0.45}$ &   $9.89^{+4.31}_{-4.77}$ & $2.51^{+0.26}_{-0.21}$ & $2.85 \pm 0.28$        & $8.48^{+0.44}_{-0.45}$ & $2.27 \pm 0.02$ & 4.27\\
20\_1   & 20 & 0.1   & $1.55^{+0.19}_{-0.52}$ &   $7.74^{+2.76}_{-3.48}$ & $2.47^{+0.29}_{-0.24}$ & $2.92^{+0.31}_{-0.30}$ & $8.24^{+0.40}_{-0.41}$ & $2.20 \pm 0.02$ & 4.75\\
\hline
30\_001 & 30 & 0.001 & $5.07^{+0.87}_{-1.25}$ & $14.25^{+5.44}_{- 6.49}$ & $1.83^{+0.56}_{-0.50}$ & $2.72^{+0.30}_{-0.28}$ & $8.82^{+0.54}_{-0.50}$ & $2.40 \pm 0.03$ & 5.52\\
30\_01  & 30 & 0.01  & $3.19^{+0.47}_{-0.65}$ & $21.17^{+6.12}_{-12.73}$ & $2.47^{+0.22}_{-0.18}$ & $2.82^{+0.28}_{-0.27}$ & $8.61^{+0.38}_{-0.41}$ & $2.32 \pm 0.03$ & 3.72\\
30\_03  & 30 & 0.03  & $2.06^{+0.24}_{-0.82}$ & $15.77^{+5.10}_{- 9.20}$ & $2.55^{+0.24}_{-0.16}$ & $2.86 \pm 0.27$        & $8.66^{+0.41}_{-0.46}$ & $2.28 \pm 0.03$ & 3.57\\
30\_05  & 30 & 0.05  & $1.52^{+0.25}_{-0.45}$ & $10.67^{+3.68}_{- 5.04}$ & $2.58^{+0.22}_{-0.19}$ & $2.86 \pm 0.28$        & $8.48^{+0.42}_{-0.44}$ & $2.26 \pm 0.02$ & 3.87\\
30\_1   & 30 & 0.1   & $1.20^{+0.15}_{-0.31}$ &  $7.97^{+2.82}_{- 3.56}$ & $2.51^{+0.28}_{-0.23}$ & $2.92 \pm 0.31$        & $8.23^{+0.39}_{-0.41}$ & $2.20 \pm 0.02$ & 4.62\\
\hline
\end{tabular}
\end{center}
\end{table*}

\section{Fitting procedure and results}
\label{Sect_Fit}

\begin{figure*}
\centering
\includegraphics[width=0.48\textwidth]{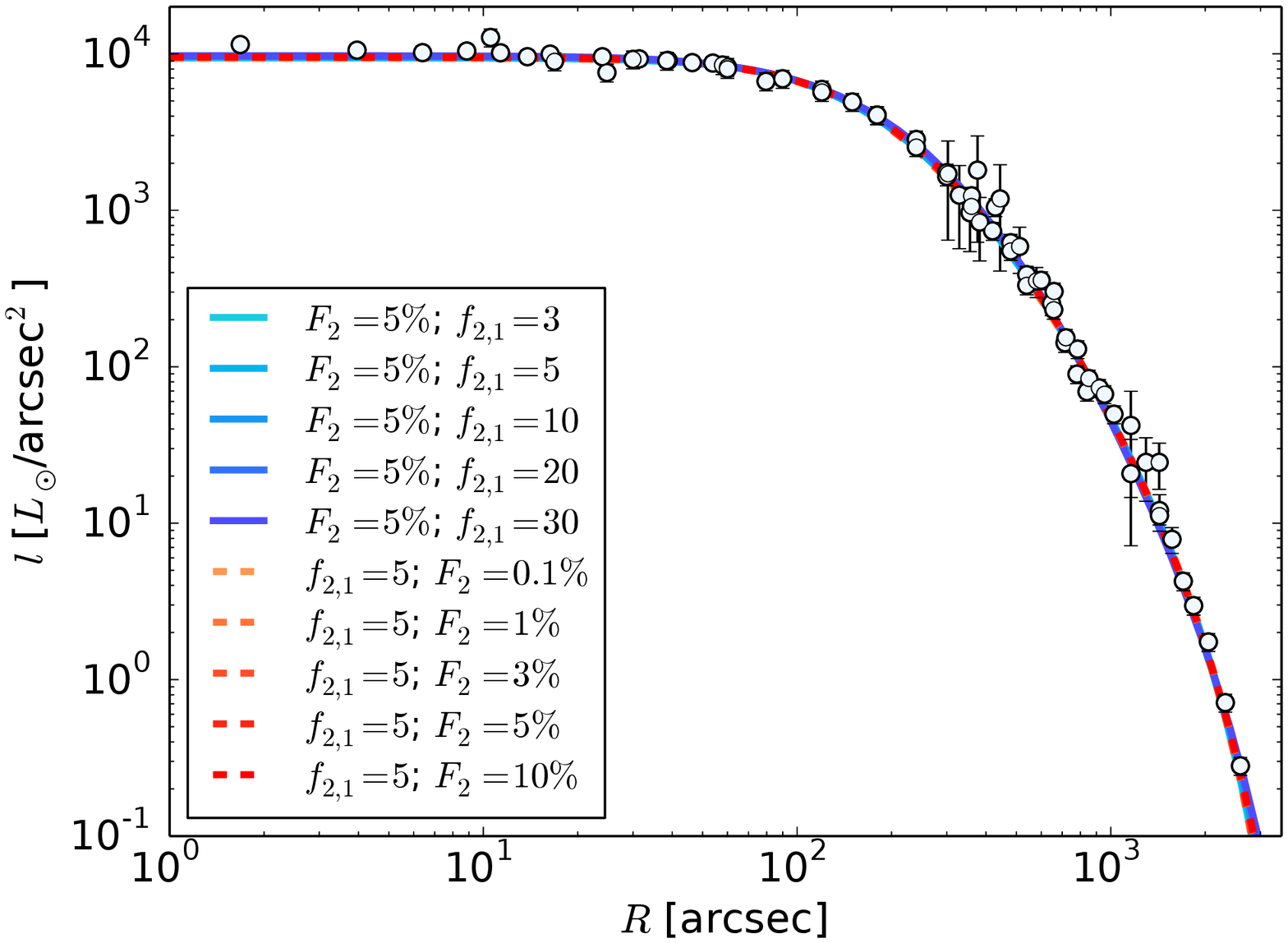} \quad
\includegraphics[width=0.48\textwidth]{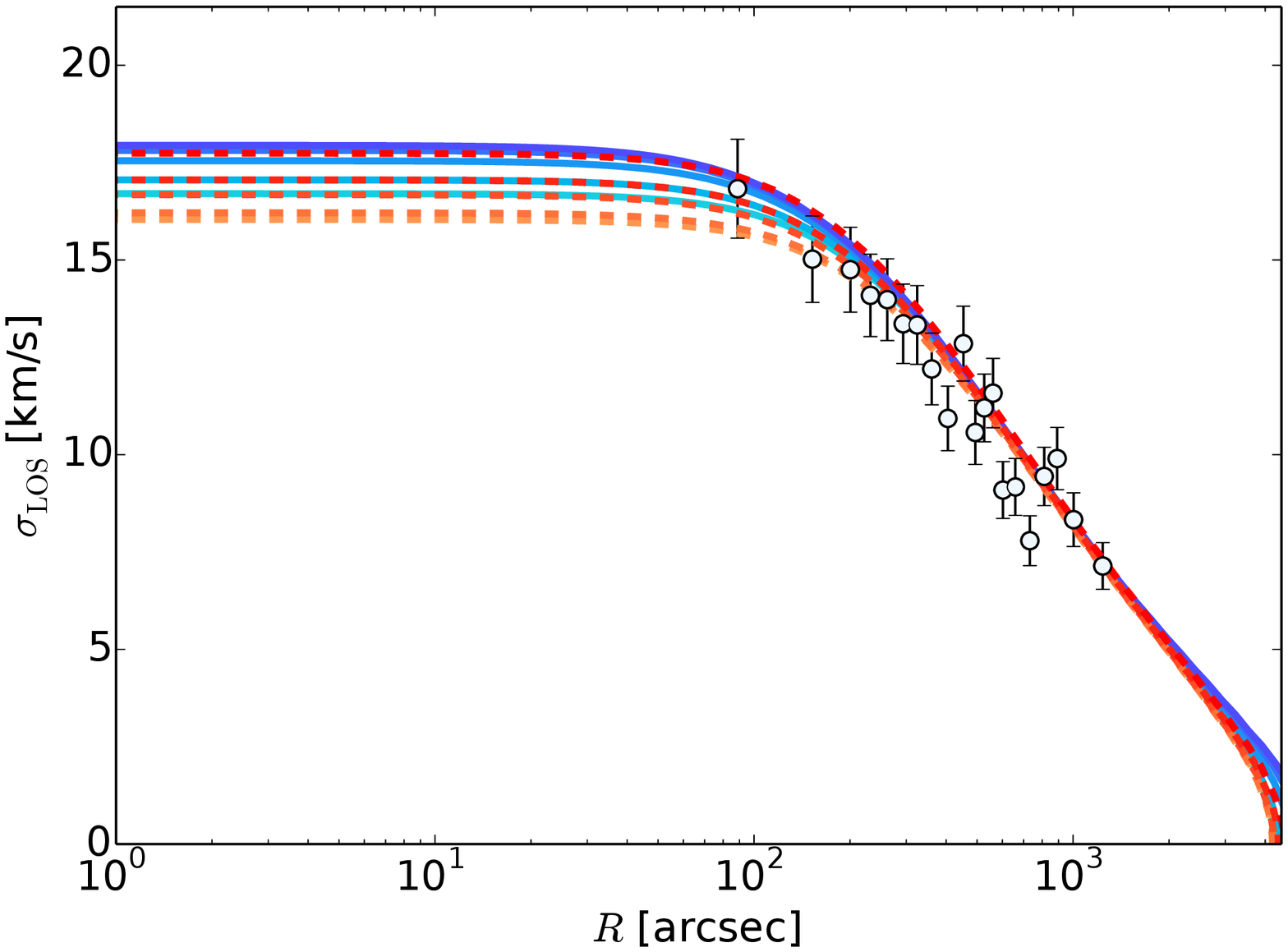} \\
\includegraphics[width=0.48\textwidth]{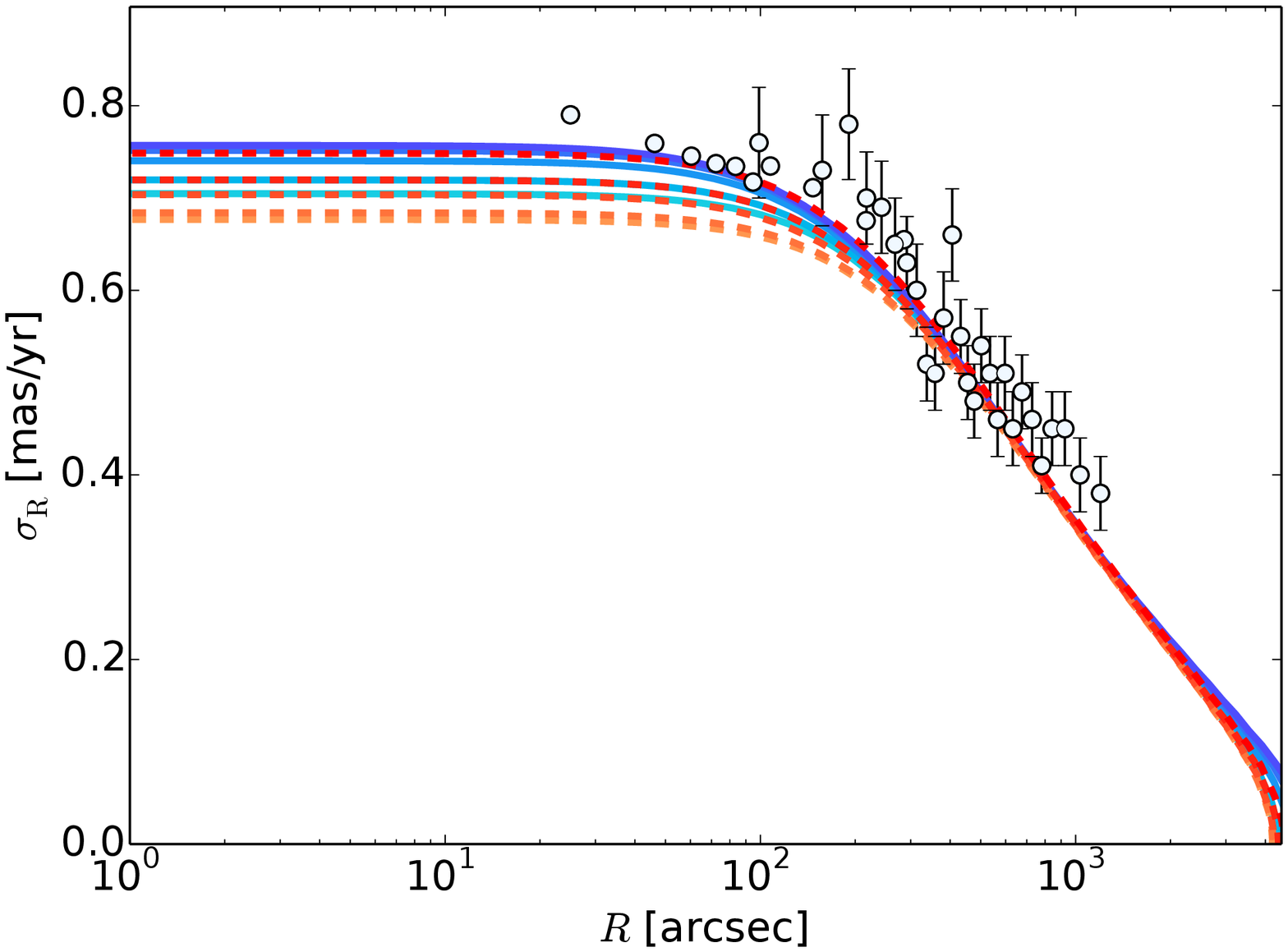} \quad
\includegraphics[width=0.48\textwidth]{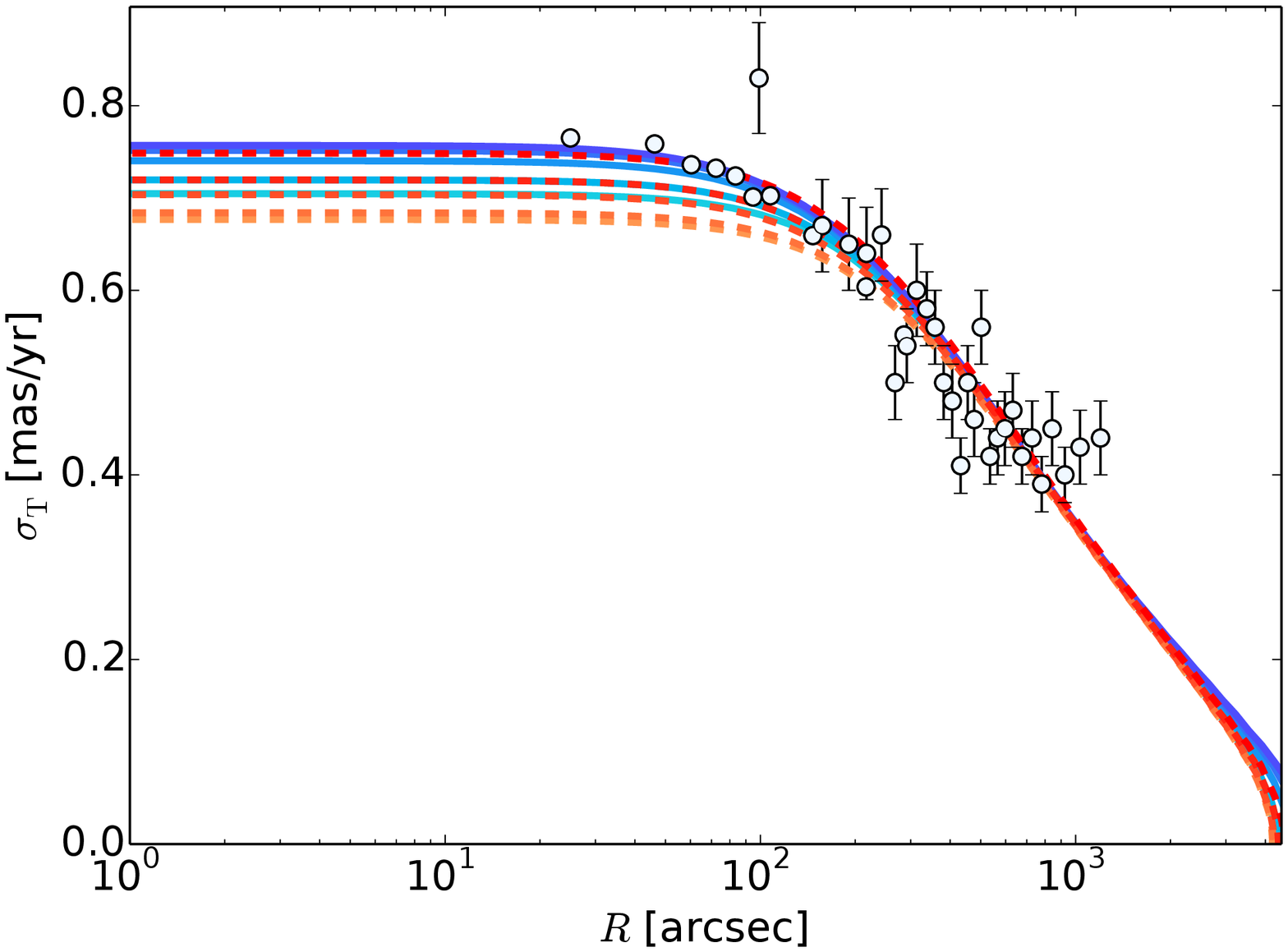} \\
\caption{From top left, proceeding clockwise: surface brightness profile, line-of-sight velocity dispersion profile, tangential and radial proper motion velocity dispersion profiles for $\omega$ Cen. Red dashed lines represent the best-fit models obtained when considering $f_{2,1} = 5$ (i.e., $m_2 = 1.5$ M$_{\odot}$), and different values of $F_2$; blue lines represent the best-fit models obtained when considering $F_2 = 5 \%$, and different values of $f_{2,1}$ (see labels and Table~\ref{Tab1}). The surface brightness profile is from \citet{TKD1995} and \citet{Noyola2008}. The line-of-sight velocity dispersion is calculated from velocities by \citet{Pancino2007} and \citet{Reijns2006}. The proper motions velocity dispersion profiles are calculated from data by \citet{vanLeeuwen2000} and \citet{AvdM2010}.}
\label{Fig_mm}
\end{figure*}

\begin{figure*}
\centering
\includegraphics[width=0.48\textwidth]{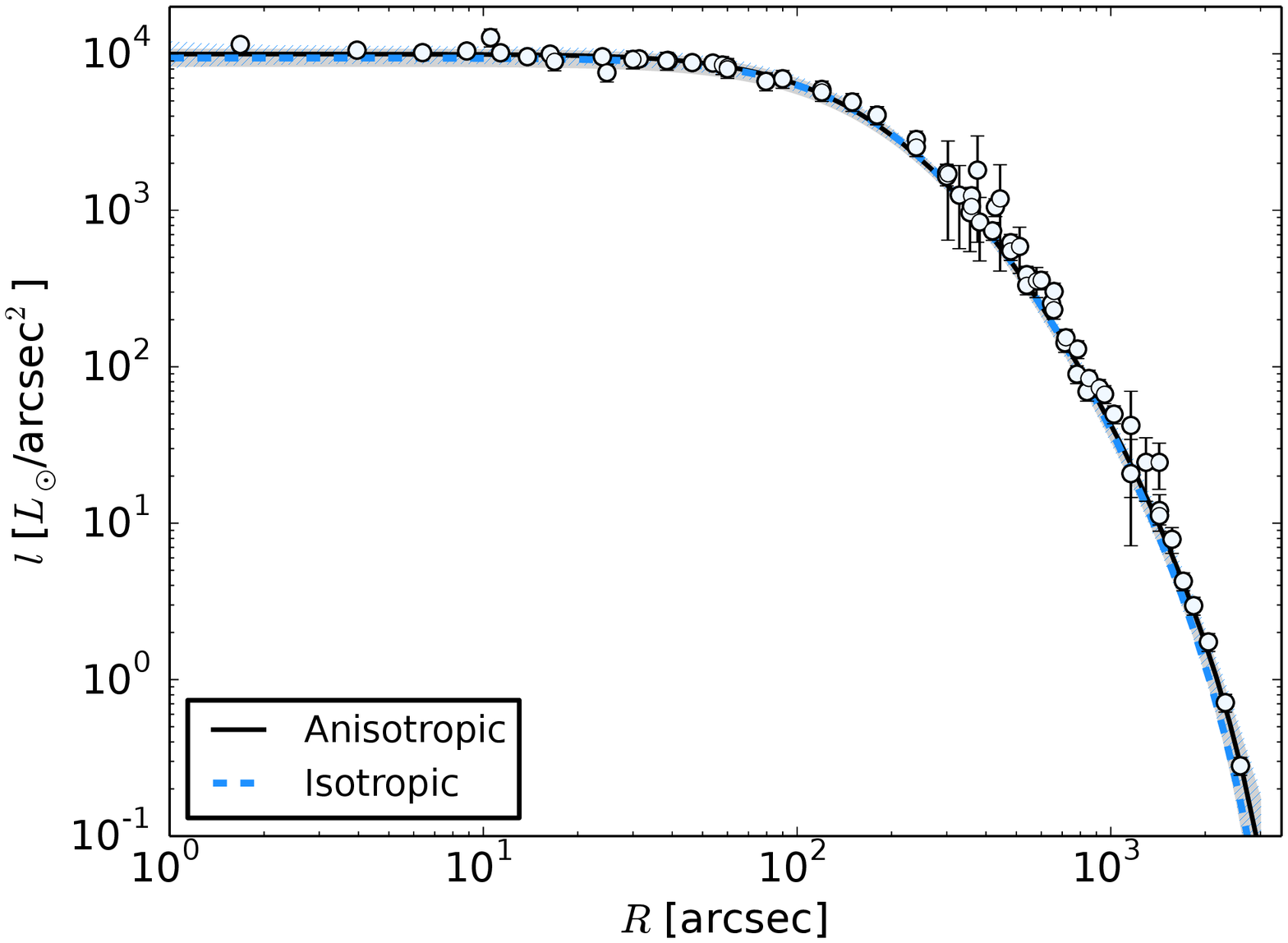} \quad
\includegraphics[width=0.48\textwidth]{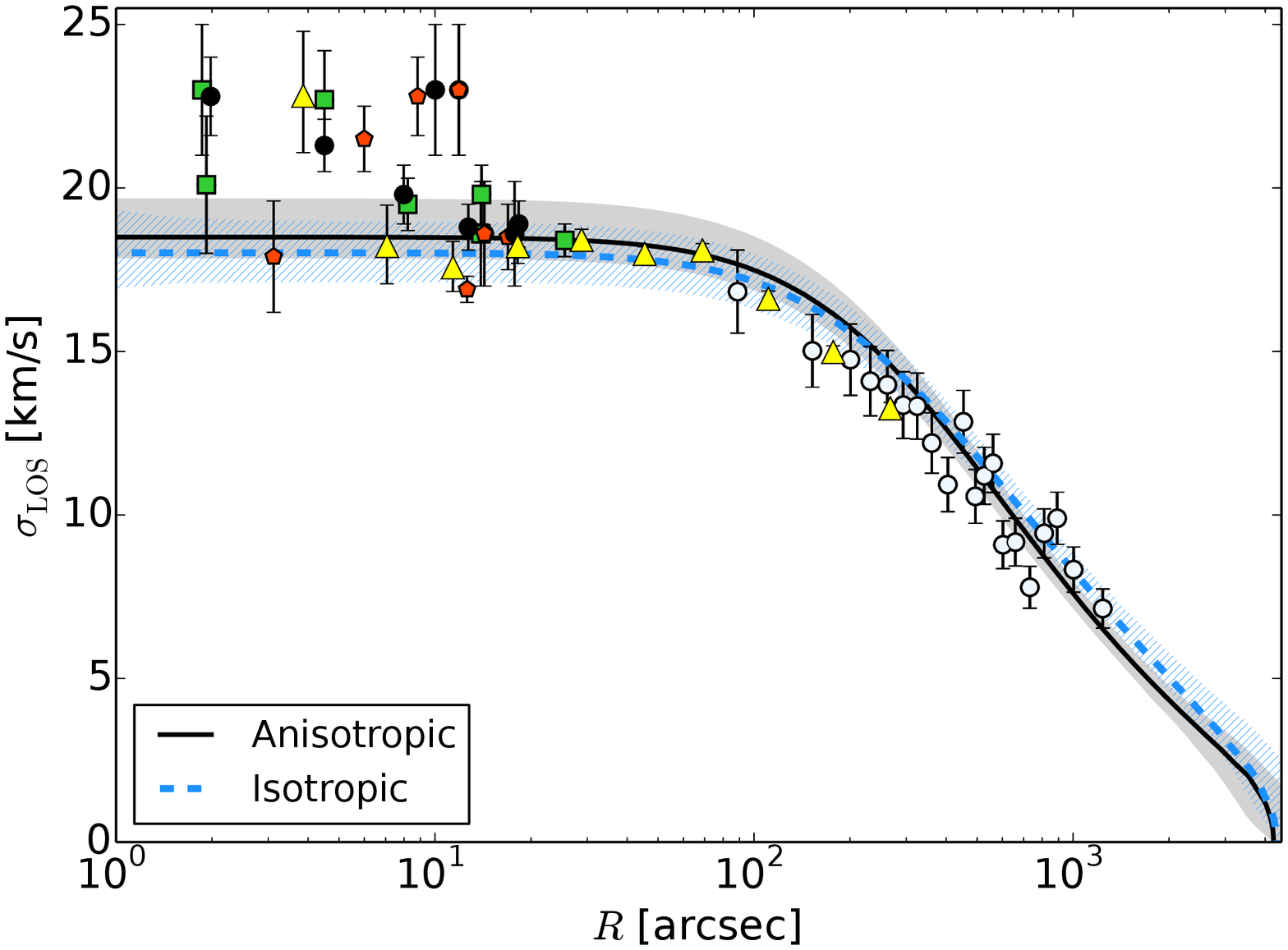} \\
\includegraphics[width=0.48\textwidth]{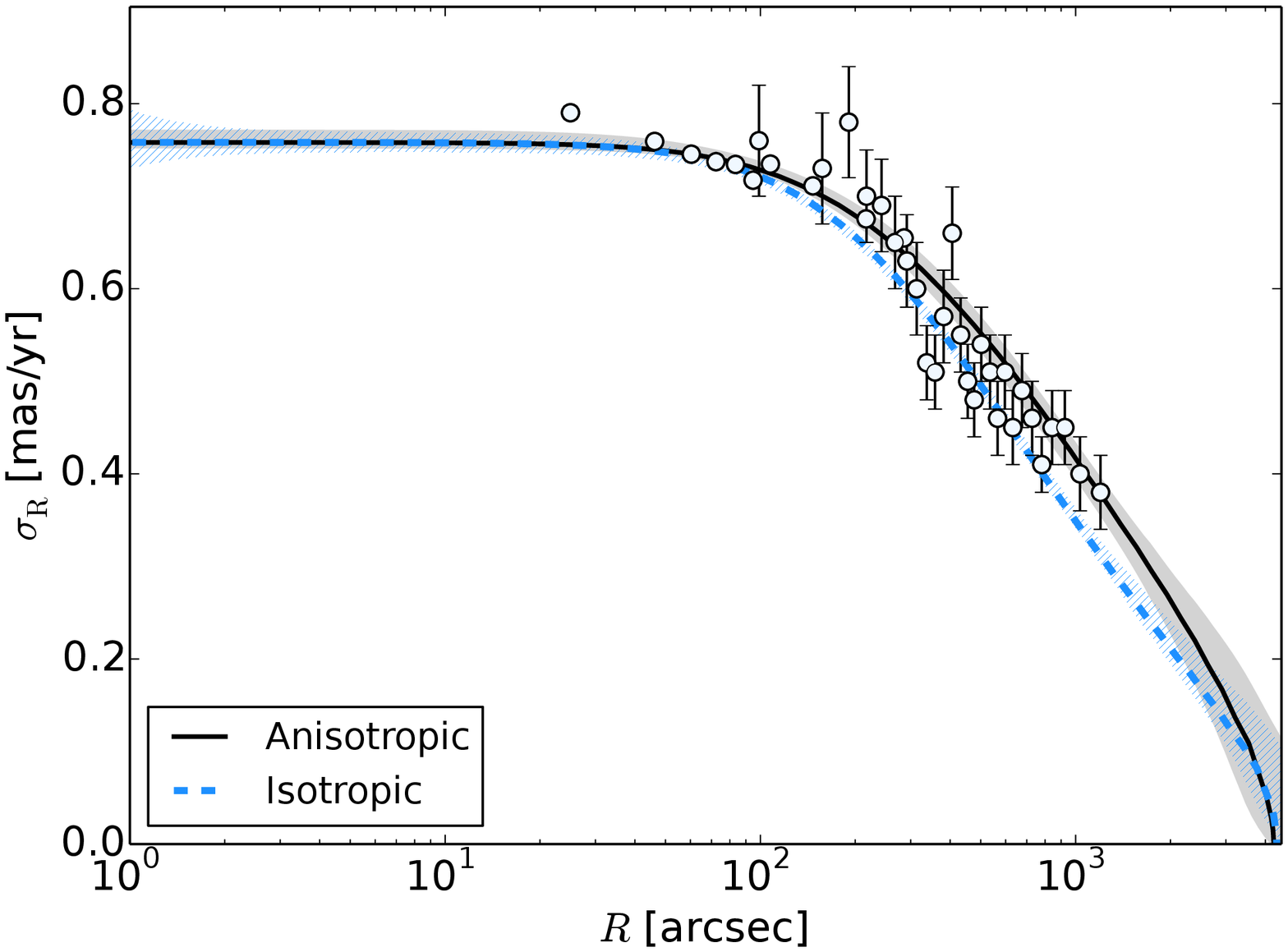} \quad
\includegraphics[width=0.48\textwidth]{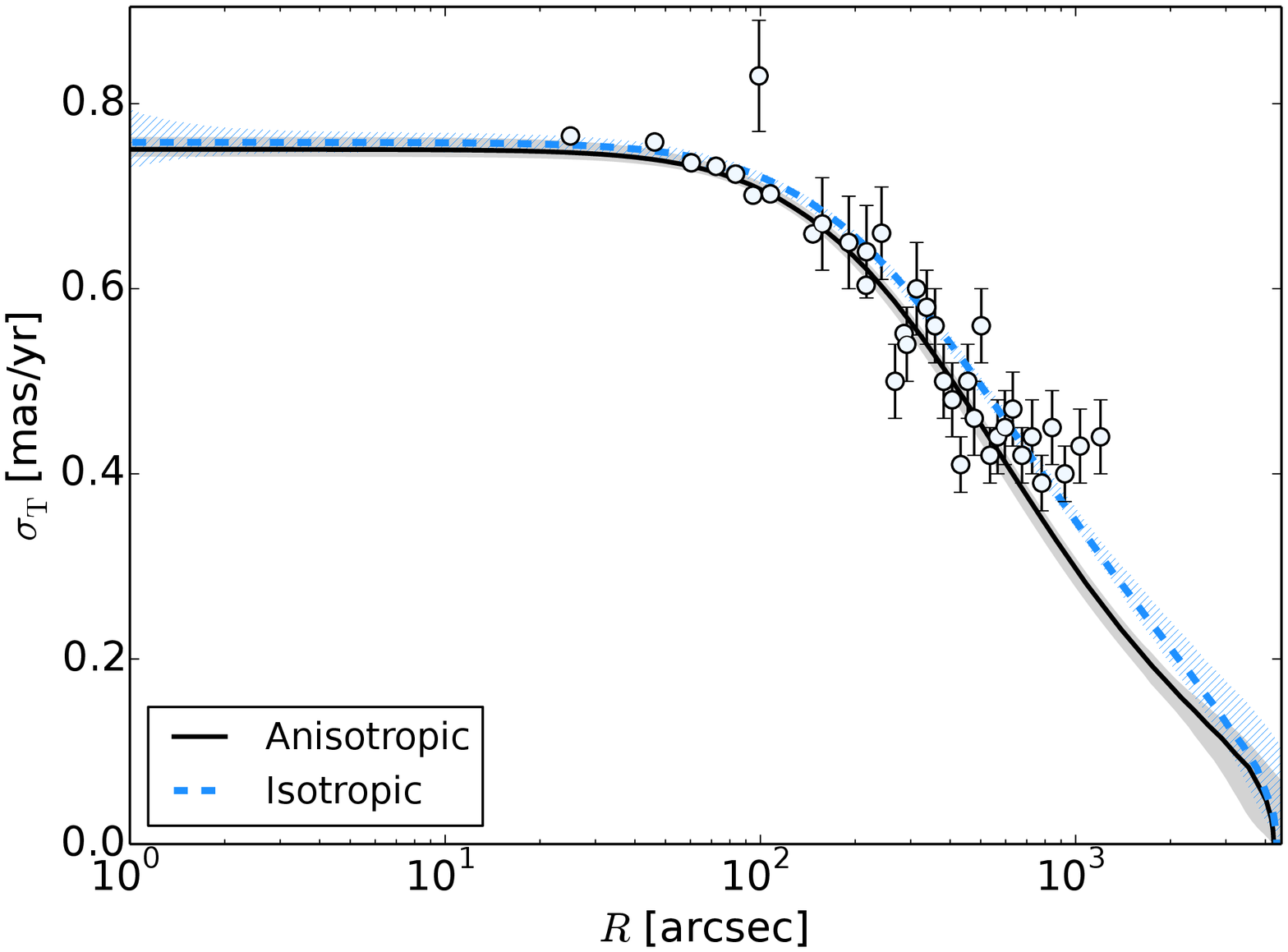} \\
\caption{From top left, proceeding clockwise: surface brightness profile, line-of-sight velocity dispersion profile, tangential and radial proper motion velocity dispersion profiles for $\omega$ Cen. Solid black lines and dashed blue lines reproduce respectively the radial profiles of the anisotropic and isotropic best-fit model obtained with the one-step fitting procedure. The shaded and dashed areas represent respectively the anisotropic and isotropic models that occupy a $1\sigma$ region around the maximum likelihood, as identified by \emcee. The data profiles are the same as in Fig.~\ref{Fig_mm}. In addition, in the top right panel, we also show the velocity dispersion profiles calculated from integrated spectra: the black dots reproduce the profile provided by \citet{Noyola2010} when considering their kinematical centre, the green squares the profile they obtained when considering the centre by \citet{Noyola2008}, and the red pentagons the one with the centre by \citet{AvdM2010}. Moreover, the yellow triangles reproduce the profile by \citet{Kamann2018}, derived from  MUSE data.}
\label{Fig_1S}
\end{figure*}

Our goal is to show what the effect is of the presence of a population of heavy remnants on the observable quantities, and in particular to show how this affects the velocity dispersion profile, which is the main observable used to infer the presence of IMBHs in globular clusters. All the observational quantities are compared with the profiles of the visible component of the models; the other component is never directly compared with data. In this Section we describe the fitting procedures we adopted.  

We first carry out a \textit{two-steps fit}: the first step involves a fit of models with a given mass function to the surface brightness profile to determine the structural parameters and the scales, and the second step a fit determining the vertical scaling of the line-of-sight velocity dispersion profile, defined by means of the mass-to-light ratio. In our previous analysis \citep{Zocchi2017} of $\omega$ Cen with anisotropic models, we decided to carry out a two-steps fitting in order to follow a procedure as close as possible to the one used in analyses based on the Jeans equation \citep[e.g., see][]{Noyola2010,vdMA2010,Watkins2013}. Indeed, with Jeans approach, the first step is a fit to the surface brightness profile, which is subsequently deprojected to obtain the 3d density distribution. Then a certain $M/L$ and an anisotropy profile are chosen, and the Jeans equation is solved to obtain a 3d velocity dispersion, that is projected to be compared to the data; this second step is repeated until the models match the observed profiles. Here we are using the same two-step fitting procedure (see Section~\ref{Sect_Res_twostep}) because it allows us to explore the dynamics of models with different amounts of BHs reproducing equally well the surface brightness profile of $\omega$ Cen, enabling thus a more direct comparison of models predicting different kinematics for the visible stars (see Section~\ref{Sect_Res_twostep}).  

In addition, we also carry out a \textit{one-step fit} to determine the best model to reproduce all the observational profiles available for this cluster: in this case, 9 fitting parameters are determined at once by fitting the models to all the profiles (see Section~\ref{Sect_Res_onestep}).

We used the surface brightness profile and the line-of-sight and proper motions velocity dispersion profiles of $\omega$ Cen. For a detailed description of these profiles, we refer the reader to Section 3 of \citet{Zocchi2017}, as we use the same set of data for the present analysis. The surface brightness profile is composed by data from \citet{TKD1995} and \citet{Noyola2008}; line-of-sight velocities of single stars are taken from \citet{Reijns2006} and \citet{Pancino2007}; ground-based proper motions are from \citet{vanLeeuwen2000}, and \textit{HST} data from \citet{AvdM2010}.

We carry out the fits by using \emcee\ \citep{emcee_paper}, a \python\ implementation of Goodman and Weare's affine invariant Markov chain Monte Carlo ensemble sampler.\footnote{\emcee\ is available at \href{https://github.com/dfm/emcee}{https://github.com/dfm/emcee}.}

\subsection{Two-steps fitting}
\label{Sect_Res_twostep}

In order to understand the effect of the presence of the dark population on the visible component, we carry out fits of models with different mass functions, by varying the values of the parameters $f_{2,1}$ and $F_2$. To do this, we select several values for $f_{2,1}$ and $F_2$ to explore a large range of possible scenarios. With the values chosen for these parameters, the mean mass of objects in the dark component is $3 \leq f_{2,1} \leq 30$ (this means that, when considering $m_1 = 0.3$ M$_{\sun}$, $m_2$ ranges from 0.9 M$_{\sun}$ to 9 M$_{\sun}$), and the total mass they account for ranges from $0.1 \%$ to $10 \%$ of the total mass of the cluster. We deliberately chose extreme values for the boundaries of the grid of values to fully probe the parameter space. The distance to $\omega$ Cen is fixed to 5 kpc, as in \citet{Zocchi2017}. In order to determine the importance of remnants independently from radial anisotropy, in this Section we consider isotropic models.

As in \citet{Zocchi2017}, we first determine the model structural parameters ($W_0$ and $g$) and the scales (the luminosity $L$, and the half-mass radius $\rh$) by fitting the models to the surface brightness profile alone: this ensures that the models we find are able to accurately represent the distribution of visible stars in the cluster. In this first step, we determine the best-fit values of the 4 fitting parameters by maximising the log-likelihood function:
\begin{equation}
\lambda \propto - \frac{\chi^2}{2} \ ,
\end{equation}
i.e., by minimising the following quantity:
\begin{equation}
 \chi_1^2 = \sum_{i = 1}^{N_{\rm SB}} \frac{\left[ l_i - \lambda_1(R_i) \right]^2 }{\delta l_i^2} \ ,
\label{eq_chi1}
\end{equation}
where $l_i$ and $\delta l_i$ are the surface luminosity and luminosity error at the radial position $R_i$ for each of the $N_{\rm SB}$ points in the surface brightness profile; $\lambda_1$ is the projected luminosity density\footnote{Because the model profiles are defined in terms of masses and not luminosities, here we consider $M/L = 1$ to perform the fit; when the best-fit value of $M/L$ is then recovered from the second step in the fitting procedure, the value of $L$ is updated accordingly.} of the visible component of the model, its shape determined by the values of all the fitting parameters. For the parameters, we adopt uniform priors over the following ranges: $1 < W_0 < 30$, $0.1 < g < 3.5$, $0.1 < L < 50$ in units of $10^6$ L$_{\odot}$, $0.1 < \rh < 50$ in units of pc. The best-fit value for each parameter is obtained as the median of the correspondent marginalised posterior probability distribution, and the errors correspond to the 16 and 84 per cent percentiles.

Then, in the second step, the mass-to-light ratio $M/L$ is obtained by determining the scaling of the line-of-sight velocity dispersion profile, by minimising the following quantity:
\begin{equation}
 \chi_2^2 = \sum_{i = 1}^{N_{\rm VD_{los}}} \frac{\left[ \sigma_{{\rm los},i} - \sigma_{\rm LOS,0} \, \hat{\sigma}_{\rm LOS}(R_i) \right]^2 }{\delta \sigma_{{\rm los},i}^2} \ ,
\label{eq_chi2}
\end{equation}
where $\sigma_{{\rm los},i}$, and $\delta \sigma_{{\rm los},i}$ are the line-of-sight velocity dispersion and its error at the position $R_i$, for each of the $N_{\rm VD_{los}}$ points in the velocity dispersion profile; $\hat{\sigma}_{\rm LOS}$ is the velocity dispersion profile of the models, projected along the line of sight and normalized with respect to its central value, and $\sigma_{\rm LOS,0}$  is the vertical scaling needed to match the model to the data, depending on the value of the mass-to-light ratio $M/L$.

The best-fit parameters obtained with the two-step fitting procedure are indicated in Table~\ref{Tab1}. Figure~\ref{Fig_mm} reproduces the surface brightness profile and the line-of-sight, radial and tangential proper motion velocity dispersion profiles of some of the considered models; each line in these figures represents a system with a certain mass function, different from the others, as indicated by the label in the top  left panel. 

In particular, two sets of models are represented here: the blue solid lines represent the best-fit models obtained when considering $F_2 = 5 \%$, and the red dashed lines represent the best-fit models obtained when considering $f_{2,1} = 5$ (i.e., $m_2 = 1.5$ M$_{\odot}$ when adopting $m_1 = 0.3$ M$_{\odot}$). The set of red lines, therefore, can be used to inspect the effect of considering different values for the fraction of total mass in the dark component, while the set of blue lines shows the effect of assuming different values for the mean mass of objects in the dark component.

All the models reproduce the observed surface brightness profile equally well. We recall that only the line-of-sight velocity dispersion profiles are involved in the fit determining the mass-to-light ratio of the cluster\footnote{We chose not to take into account the proper motions velocity dispersion profiles in this step because in order to express them in units of km\,s$^{-1}$ it is necessary to assume a distance to the cluster; of course, with different values for the distance, different values would be obtained for the proper motion velocity dispersion in km\,s$^{-1}$, and this would in turn propagate to the best-fit value of $M/L$. Avoiding to use proper motions data to determine $M/L$ thus limits the impact of our arbitrary choice of distance on the model parameters, allowing us to isolate the effect of $F_2$ and $f_{2,1}$ on the velocity dispersion profile. In Section~\ref{Sect_Res_onestep} the distance to the cluster is a fitting parameter, and the proper motions are therefore properly taken into account.}. The kinematic profiles are always better described by models including a large fraction of dark remnants with large mass (darkest blue and red lines in the plot), because of their larger velocity dispersion in the centre. We point out that Fig.~\ref{Fig_mm} shows that models with a very different mass function reproduce the surface brightness profile of the cluster equally well but are different from each other when considering the velocity dispersion profiles: this is a clear indication of the need of kinematical data in order to truly assess the dynamics of these stellar systems and their composition.

\subsection{One-step fitting}
\label{Sect_Res_onestep}

In this Section we include the parameters defining the mass function of the model ($f_{2,1}$ and $F_2$) in a fit to all the observational profiles at the same time. Indeed, while the two-step procedure allows us to clearly illustrate variations in the kinematics caused by changes in the mean and total masses of BHs, this one-step procedure gives the values of the parameters that best describe all the profiles considered simultaneously. Moreover, we also fit on the distance $d$, and on the dimensionless anisotropy radius $\hat{r}_{\rm a}$, which sets the amount of anisotropy of the model, allowing us to explore possible degeneracy between radial anisotropy and stellar-mass BHs. We have a total of 9 fitting parameters, and we determine their best-fit values by minimising the following quantity:
\begin{equation}
 \chi^2 = \chi_{\rm SB}^2 + \chi_{\rm LOS}^2 + \chi_{\rm PM,R}^2 + \chi_{\rm PM,T}^2
\label{chi_all}
\end{equation}
where each of the terms on the right end side is in the form:
\begin{equation}
 \chi_{\rm X}^2 = \sum_{i = 1}^{N_{\rm X}} \frac{\left[ x_i -  X_1(R_{i}) \right]^2 }{\delta x_i^2} \ ,
\label{chi_all_1}
\end{equation}
where $x_i$ and $\delta x_i$ represent the observational quantity and its error at the radial position $R_i$, and $X_1$ is the corresponding model profile for the visible component. The four terms appearing in the right-hand side of equation~(\ref{chi_all}) correspond to the surface brightness profile, line-of-sight velocity dispersion profile, and radial and tangential proper motion velocity dispersion profiles, respectively. 

For the 9 fitting parameters, we adopt uniform priors over the following ranges: $1 < W_0 < 30$, $0.1 < g < 3.5$, $0.1 < M < 50$ in units of $10^6$ M$_{\odot}$, $0.1 < \rh < 50$ in units of pc, $-3 < \log F_2 < -0.9$, $2 < f_{2,1} < 15$, $1 < M/L < 5$ in solar units, $1 < d < 10$ in units of kpc, $-4 < \log \hat{r}_{\rm a} < 1.3$ (we consider $\log \hat{r}_{\rm a}$ and $\log F_2$ as fitting parameters instead of $\hat{r}_{\rm a}$ and $F_2$ to have an uninformative prior, because they can span several orders of magnitude).

We also carry out a fit to determine what is the best isotropic model to fit the data. In this case we only have 8 parameters ($\hat{r}_{\rm a}$ is not required). 

\begin{table}
\begin{center}
\caption[Best-fit parameters for $\omega$ Cen.]{Best-fit parameters for $\omega$ Cen, obtained with the one-step fitting procedure. The columns refer to the anisotropic and isotropic models. For each model, we provide the values of the concentration parameter $W_0$ and of its alternative definition $W_0^*$, the truncation parameter $g$, the mass $M$ in units of $10^6$ M$_{\sun}$, the half-mass radius $\rh$ in units of pc, the fractional mass of the cluster contained in the dark component $F_2$, the fractional mean mass of the dark to visible components $f_{2,1}$, the mass-to-light ratio $M/L$ in solar units, the distance $d$ in kpc, the dimensionless anisotropy radius $\hat{r}_{\rm a}$, and the anisotropy parameter $\kappa$. The uncertainties are indicated for all the fitting parameters. The last four lines of the table list the values of the reduced chi-squared for the various parts of the fit introduced in equations~(\ref{chi_all}) and (\ref{chi_all_1}).} 
\label{Tab2}
\begin{tabular}{ccc}
\hline\hline
             & Anisotropic            & Isotropic  \\
\hline
$W_0$        & $3.82^{+0.80}_{-1.05}$ & $3.64^{+0.74}_{-0.72}$ \\
$W_0^*$      & $9.55^{+4.10}_{-3.23}$ & $8.66^{+2.81}_{-3.40}$ \\
$g$          & $1.77^{+0.35}_{-0.34}$ & $2.20^{+0.33}_{-0.26}$ \\
$M$          & $3.01^{+0.45}_{-0.39}$ & $2.91^{+0.44}_{-0.40}$ \\
$\rh$        & $8.57^{+0.61}_{-0.62}$ & $8.28^{+0.62}_{-0.63}$ \\
$F_2$        & $ 0.045^{+0.040}_{-0.025}$ & $0.052^{+0.029}_{-0.031}$ \\
$f_{2,1}$    & $6.90^{+4.12}_{-3.02}$ & $7.92^{+1.45}_{-2.12}$ \\
$M/L$        & $2.55^{+0.35}_{-0.28}$ & $2.59^{+0.34}_{-0.27}$ \\
$d$          & $5.14^{+0.25}_{-0.24}$ & $5.01^{+0.24}_{-0.24}$ \\
$\hat{r}_{\rm a}$ & $5.97^{+1.53}_{-1.63}$ & --- \\
$\kappa$     & $1.13 \pm 0.04$ & 1.00 \\
$\widetilde{\chi}^2_{\rm SB}$  & 2.05 & 8.74 \\
$\widetilde{\chi}^2_{\rm LOS}$ & 1.67 & 1.78 \\
$\widetilde{\chi}^2_{\rm PM,R}$  & 2.25 & 12.13 \\
$\widetilde{\chi}^2_{\rm PM,T}$  & 4.00 & 5.49 \\
\hline
\end{tabular}
\end{center}
\end{table}

Figure~\ref{Fig_1S} shows the surface brightness, line-of-sight velocity dispersion, and radial and tangential proper motion velocity dispersion profiles of $\omega$ Cen. In these plots, the black solid lines represent the anisotropic best fit model, the dashed blue lines the isotropic one. The best-fit parameters obtained with the one-step fitting procedure for the isotropic and anisotropic models are indicated in Table~\ref{Tab2}. In the table we also indicate the errors on the parameters, which in some cases appear to be quite large: this is probably due to the large number of parameters we consider in this fit, and to the large errors on the data. Another reason for the large errors is the fact that we consider as fitting parameters $f_{2,1}$ and $F_2$: because they have a very important role in determining the shape of the model profiles, by changing them slightly a large spread is obtained in the values of $W_0$ in order to reproduce the observed profile.

To illustrate how different the definition of the concentration parameter is in the case of multiple components with respect to the single component case, we also include in Table~\ref{Tab2} (as well as in Table~\ref{Tab1}) the values of this parameter obtained when considering a different definition for the mean mass of the stars in the system (see Section~\ref{Sect_Models_Params}). By inspecting the tables, it is clear that the best-fit values of $W_0$ are smaller than the ones generally obtained with King models for this cluster, while the values of its alternative definition, $W_0^*$, are much larger. When considering \citet{King1966} models, large values of concentration indicate a core-collapsed cluster, but for the multimass models we consider here this is not true, and in general it is not trivial to have a general criterion to indicate whether a cluster is core-collapsed or not.

We notice that the best-fit values of the total mass and of the half-mass radius obtained with these two models are consistent with each other (and with those of best-fit models obtained with the two-step procedure). The distance obtained for the isotropic model is very close to the one assumed in the two-step fitting procedure, while the one inferred with the anisotropic model is a bit larger, but they are all consistent within 1$\sigma$.

\begin{figure}
\centering
\includegraphics[width=0.48\textwidth]{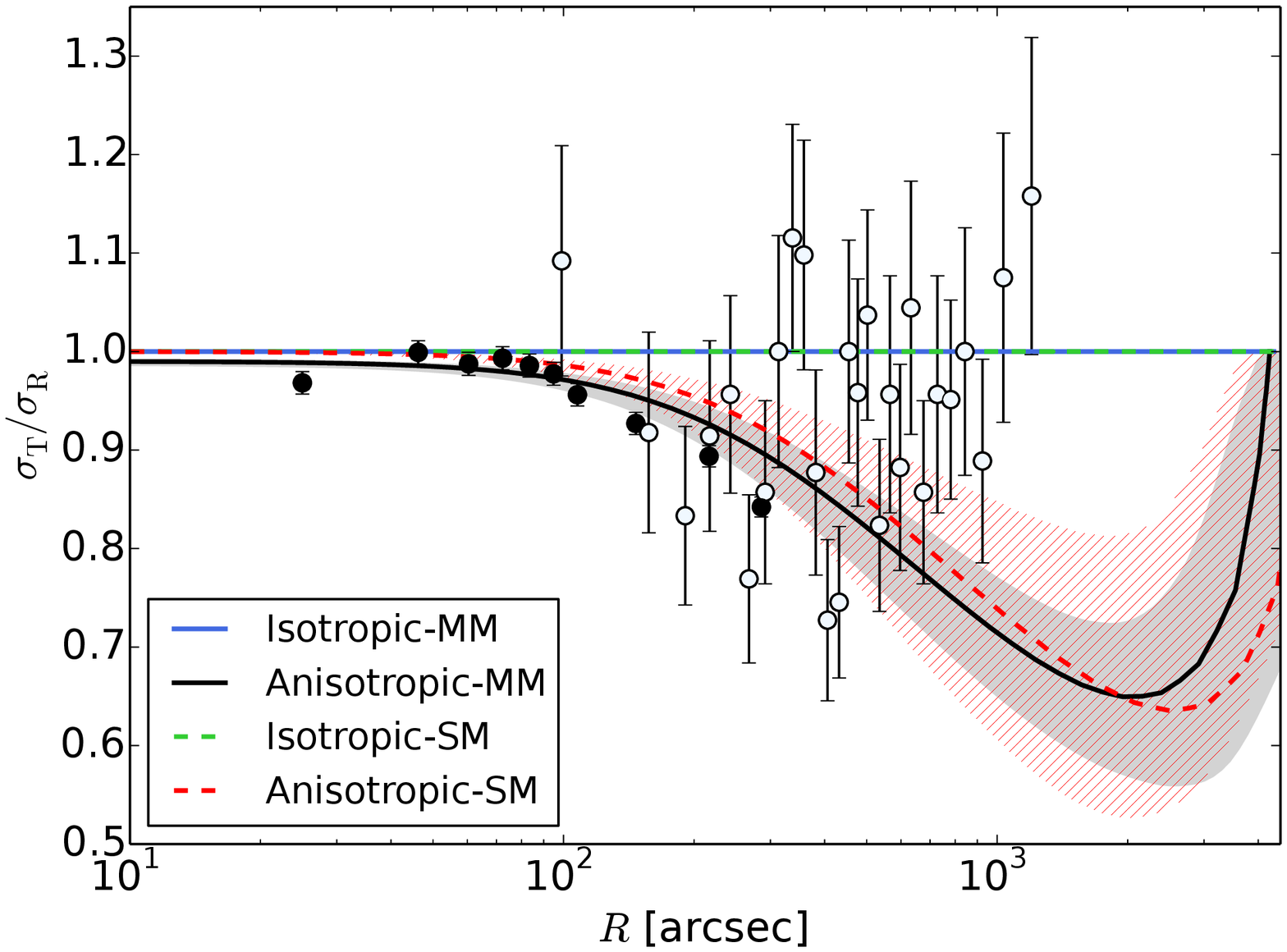} \\
\caption{Anisotropy profile $\sigma_{\rm T}/\sigma_{\rm R}$ of $\omega$ Cen, calculated as the ratio of the tangential to radial components of the proper motions velocity dispersion profiles. Black dots represent the anisotropy calculated by \textit{HST} data from \citet{AvdM2010}, white circles the anisotropy calculated by ground-based data from \citet{vanLeeuwen2000}. The dashed green and red lines indicate the isotropic and anisotropic single-mass best-fit models obtained by \citet{Zocchi2017}; the solid blue and black lines represent the isotropic and anisotropic multimass best-fit models described in Section~\ref{Sect_Res_onestep}. The shaded and dashed areas represent respectively the anisotropic multimass and single-mass models that occupy a $1\sigma$ region around the maximum likelihood, as identified by \emcee.}
\label{Fig_beta_cfr}
\end{figure}

The best-fit anisotropic model reproduces the observed profiles better than the isotropic one, suggesting that, as expected, anisotropic models are more indicated to describe $\omega$ Cen. This conclusion is supported by Fig.~\ref{Fig_beta_cfr}, which reproduces the anisotropy profiles (calculated as the ratio of tangential to radial proper motions velocity dispersion profiles, $\sigma_{\rm T}/\sigma_{\rm R}$: $\sigma_{\rm T}/\sigma_{\rm R} < 1$ indicates radial anisotropy, $\sigma_{\rm T}/\sigma_{\rm R} > 1$ indicates tangential anisotropy, and $\sigma_{\rm T}/\sigma_{\rm R} = 1$ indicates isotropy) for the best-fit models obtained with the one-step fitting procedure in this paper for multimass models and in \citet{Zocchi2017} for single-mass models. In this plot, the dashed green and solid blue lines indicate the single-mass and multimass isotropic best-fit models, and the dashed red and solid black lines the anisotropic single-mass and multimass best-fit models. Anisotropic models better reproduce the shape of the observed profile (especially the \textit{HST} data) with respect to isotropic models. A discrepancy  is seen between the models and the ground-based anisotropy data at large distances from the centre, but the large scatter of these data prevents us from drawing any firm conclusion from this. With the forthcoming Gaia proper motions data it will be possible to extend the profile currently available to even larger radii, up to the edge of the cluster, and to investigate this further. If the tendency towards tangential anisotropy will be confirmed, different models will need to be considered in order to reproduce the profile \citep[such as for example those by][]{VarriBertin2012}.

For the anisotropic best-fit model, we compute the anisotropy profiles for the two components, to understand how radial anisotropy is distributed in each component. The anisotropy profile for the dark component has a deeper minimum (i.e., it is more radially anisotropic), $\min(\sigma_{\rm T}/\sigma_{\rm R}) \sim 0.4$, than the visible component, $\min(\sigma_{\rm T}/\sigma_{\rm R}) \sim 0.7$ (see black line in Fig.~\ref{Fig_beta_cfr}). However, the BHs are more concentrated in the innermost part of the cluster, in a region where these profiles are similar to each other, and where their value is $\sigma_{\rm T}/\sigma_{\rm R} \sim 1$, which indicates isotropy. Therefore, the invisible component is not characterised by strong radial anisotropy. This is confirmed by computing the anisotropy parameter $\kappa = 2 K_{\rm r}/K_{\rm t}$, defined as the ratio between the radial and tangential components of the kinetic energy \citep{PolyachenkoShukhman1981,FridmanPolyachenko1984}: $\kappa > 1$ for radially anisotropic systems, $\kappa < 1$ for tangentially anisotropic systems, and $\kappa = 1$ for isotropic systems. The value of this parameter for the system as a whole is given in Table~\ref{Tab2}; when considering each component separately, we obtain $\kappa_1 = 1.12$, and $\kappa_2 = 1.06$, confirming that the light stars are more anisotropic.

In the top right panel of Fig.~\ref{Fig_1S} we also show the velocity dispersion profile calculated from integrated spectra obtained by \citet{Noyola2010} in the inner region of the cluster. The black dots reproduce the profile provided by \citet{Noyola2010} when considering their kinematical centre, the green square the profile obtained when considering the centre by \citet{Noyola2008}, and the red pentagons the one with the centre by \citet{AvdM2010}. In addition, we overplot the velocity dispersion profile recently calculated from data obtained with MUSE observations by \citet{Kamann2018}; this profile is represented here with yellow triangles. By inspecting the figure we see that, even though the best-fit models have been calculated without using these data, the best-fit two-components models obtained here can partially reproduce their behaviour; in particular, the \citet{Kamann2018} profile is well represented by our best-fit models, with the only exception of the innermost data point, which is $\sim 2 \sigma$ away from our models. We additionally carried out the fitting procedure by using also these data, and the resulting best-fit model is very similar to the one obtained without taking these data into account. We refer the reader to Section~5.2.1 of \citet{Zocchi2017} for a comparison to the results obtained with anisotropic single-mass models.


\begin{figure}
\centering
\includegraphics[width=0.48\textwidth]{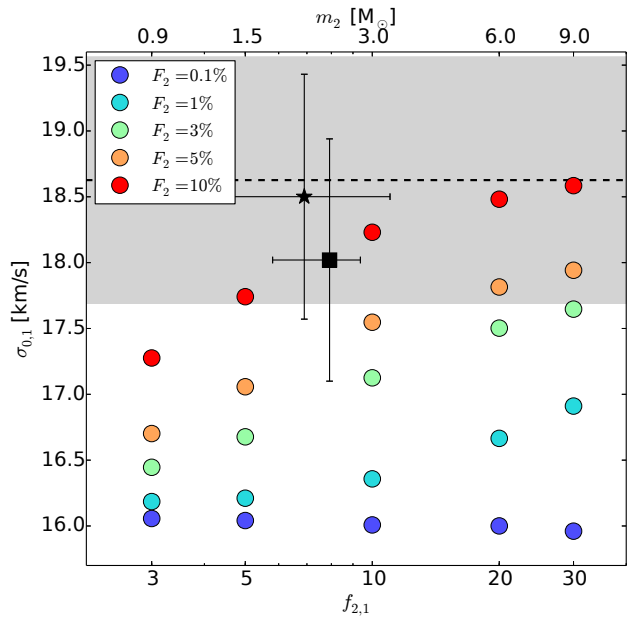}
\caption{Central line-of-sight velocity dispersion as a function of the fraction of mean mass of stars in the dark component to mean mass of stars in the visible component, $f_{2,1}$; the mean mass of stars in the dark component, $m_2$ is also indicated in the top, as obtained when assuming $m_1 = 0.3$ M$_{\sun}$. Coloured circles correspond to the values found for the two-steps fitting procedure, for the isotropic models listed in Table~\ref{Tab1}; each colour indicates a different value of the fraction of the total mass of the cluster contained in the dark component, $F_2$, as indicated by the label. The black square and the black star correspond to the values found for the one-step fit described in Section~\ref{Sect_Res_onestep} for isotropic and anisotropic models respectively. The dashed line marks the value of $\sigma_{0}$ obtained in \citet{Zocchi2017}, and the shaded grey area represents the error on this value.}
\label{Fig_sigma0_mBH}
\end{figure}

\section{Discussion}
\label{Sect_Res}

\subsection{Central line-of-sight velocity dispersion}

Among the observational features that can be linked to the presence of an intermediate-mass black hole in the centre of globular clusters, one of the most widely sought-after is a rise in the line-of-sight velocity dispersion profile towards the centre. \citet{Zocchi2017} showed that the increase of the velocity dispersion towards the centre of the cluster can be partially accounted for by the presence of radial anisotropy in the system, therefore limiting the room for the presence of a massive IMBH. 

\begin{table*}
\begin{center}
\caption[Comparison with literature.]{Comparison with literature. For each of the previous works presenting a dynamical study of $\omega$ Cen, we list the reference, the values of mass $M$, mass-to-light ratio $M/L$ and distance $d$ obtained therein, and a brief description of the models used. Values between square brackets are fixed beforehand, and do not result from a fitting procedure.} 
\label{Tab_Cfr_Literature}
\begin{tabular}{ccccccccccc}
\hline\hline
Reference  & $M$ & $M/L$ & $d$ & Models \\
  & [$10^6$ M$_{\odot}$] & [M$_{\odot}$/L$_{\odot}$] & [kpc] & \\
\hline
\citet{Meylan1987}   & 3.9                        & 2.9                     & [5.2]                   & multimass anisotropic \citet{Michie1963} models \\
\citet{Meylan1995}   & 5.1                        & 4.1                     & [5.2]                   & multimass anisotropic \citet{Michie1963} models \\
\citet{vandeVen2006} & 2.5 $\pm$ 0.3              & 2.5 $\pm$ 0.1           & 4.8 $\pm$ 0.3           & axisymmetric rotating orbit-based models \\
\citet{vdMA2010}     & 2.8                        & 2.62 $\pm$ 0.06         & 4.73 $\pm$ 0.0          & anisotropic models (Jeans) \\
\citet{Watkins2013}  &                            & 2.71 $\pm$ 0.05         & 4.59 $\pm$ 0.08         & anisotropic models (Jeans) \\
\citet{BVBZ2013}     & 1.953 $\pm$ 0.16           & 2.86 $\pm$ 0.14         & 4.11 $\pm$ 0.07         & rotating models \citep{VarriBertin2012} \\
\citet{Watkins2015}  & 3.452 $^{+0.145}_{-0.143}$ & 2.66 $\pm$ 0.04         & 5.19 $^{+0.07}_{-0.08}$ & isotropic models (Jeans) \\
\citet{DeVitaBZ2016} & 3.116                      & 2.87                    & [5.2]                   & anisotropic $f^{(\nu)}_{\rm T}$ models \\
\citet{DeVitaBZ2016} & 3.02                       & 2.04                    & [5.2]                   & anisotropic $f^{(\nu)}_{\rm T}$ models with two mass components \\
\citet{Baumgardt2017}& 2.95 $\pm$ 0.02            & 2.54 $\pm$ 0.26         & 5.00 $\pm$ 0.05         & $N$-body simulations \\
\citet{Zocchi2017}   & 3.24 $^{+0.51}_{-0.47}$    & 2.92 $^{+0.36}_{-0.32}$ & 5.13 $\pm$ 0.25         & anisotropic \limepy\ models \\
\hline
this work            & 3.01 $^{+0.45}_{-0.39}$    & 2.55 $^{+0.35}_{-0.28}$ & 5.14 $^{+0.25}_{-0.24}$ & anisotropic \limepy\ models with two mass components \\
\hline
\end{tabular}
\end{center}
\end{table*}

Here we explore the effect of the presence of a centrally concentrated population of dark remnants on this observable. In globular clusters, two-body encounters bring the systems towards a state of partial energy equipartition \citep{Spitzer1987,TrentivdM2013,Bianchini2016,Peuten2017}: massive stars tend to lose kinetic energy in the encounters, and sink in the centre of the cluster, while low-mass stars gain kinetic energy and move towards the outer parts. This process induces mass segregation and causes the system to have mass-dependent kinematics, with fast low-mass stars and slow massive stars. Therefore, in a globular cluster containing a large number of remnants we expect to observe a larger velocity dispersion for visible stars, with respect to the one we would measure in a cluster in which the red giant stars are the most massive ones. We investigate the magnitude of this effect by using the results of the fits on $\omega$ Cen.

Figure~\ref{Fig_sigma0_mBH} shows the values of the central line-of-sight velocity dispersion $\sigma_{0,1}$ obtained for the visible component for each model as a function of the mean mass of BHs. Coloured circles correspond to the values found for the isotropic models obtained with the two-steps fitting procedure described in Section~\ref{Sect_Res_twostep} (see also Table~\ref{Tab1} for a full list); each colour indicates the value of the fraction of the total mass of the cluster in the dark component, $F_{2}$, as indicated in the legend on the top left. The main result observable in the plot is that, as expected, by increasing the fraction of remnants in the cluster, the value of $\sigma_{0,1}$ increases: this is seen by looking at points from the bottom to the top, for each value of $f_{2,1}$. We notice that the quality of the fit, as indicated by the corresponding values of $\widetilde{\chi}^2$ in Table~\ref{Tab1}, is better for models with larger values of $F_{2}$. The value of the central line-of-sight velocity dispersion for the visible component increases also, for a constant value of $F_2$, when increasing $f_{2,1}$. We notice that this trend is not true for the case with $F_2 = 0.1 \%$: in this case $\sigma_{0,1}$ decreases when increasing $f_{2,1}$, possibly because of the constraint induced by the fitting procedure, which requires the models to accurately reproduce the surface brightness profile for each choice of the mass function we adopted.

In Fig.~\ref{Fig_sigma0_mBH} we also show the values found for the one-step fit (see Section~\ref{Sect_Res_onestep}) for isotropic and anisotropic models, indicated with the black square and the black star, respectively. The dashed line marks the value of $\sigma_{0}$ obtained in \citet{Zocchi2017}, and the shaded grey area represents the error on this value. From the plot we see that the largest value of $\sigma_{0,1}$ is obtained, for isotropic models, when considering the presence of a very large population of massive dark remnants ($F_2 = 10\%$ and $f_{2,1} = 30$, i.e. $m_2 = 9$ M$_{\sun}$ when assuming $m_1 = 0.3$ M$_{\sun}$). \citet{Zocchi2017} showed that a similar value is reached for the most radially anisotropic model without mass segregation they consider ($\kappa = 1.3$, see their Fig. 7). However, when adopting the one-step fitting procedure and allowing the models to be both radially anisotropic and mass segregated it is possible to obtain the same value of $\sigma_{0,1}\sim18.6$ km~s$^{-1}$ with a smaller population of remnants ($F_2 = 4.5 \%$), and with a less extreme degree of anisotropy ($\kappa = 1.13$).

\subsection{Comparison with previous works}

In Table~\ref{Tab_Cfr_Literature} we list the values obtained with different models for the mass, mass-to-light ratio, and distance of $\omega$ Cen; we compare the results in the literature with those we present here. The values of the mass and mass-to-light ratio we obtain for $\omega$ Cen are compatible within $2\sigma$ with all the previous estimates, except for those proposed by \citet{Meylan1995}. As for the distance, our estimate is larger than those by \citet{Watkins2013} and \citet{BVBZ2013}, but it is compatible with all the others within $2\sigma$. 

The only other work in the literature adopting dynamical models with two mass components (one visible, describing stars, and the other invisible, describing remnants) to reproduce this cluster is the one by \citet{DeVitaBZ2016}. The authors impose a value of $f_{2,1} = 3$ and $F_2 = 0.33$ to set up the models, and they find a best-fit total mass for invisible component of $7.5 \times 10^5$ M$_{\sun}$. In the present work we find a mass of $1.35 \times 10^5$ M$_{\sun}$ for the invisible component ($F_2 \sim 4.5\%$). Even though the mass function of these works is different, the total mass obtained for the cluster is remarkably similar, and the value of the mass-to-light ratio is compatible within $2\sigma$. An additional comparison is possible with the work presented by \citet{DeVitaBZ2016}, as they present the mass-to-light ratio radial profile of their best-fit models (see the right panel of their Fig.~12). The profile they obtained for models having an invisible heavy component decreases from $M/L \sim 2.8$ in the centre to $M/L \gtrsim 2$ at about 30 pc from the centre. The mass-to-light ratio profile of our anisotropic best-fit model is also decreasing, but more steeply: we find $M/L \sim 4.5$ in the centre and $M/L \sim 2.5$ at about 7 pc from the centre; interestingly, we find the same behaviour also for our isotropic best-fit model. The difference in the values of $M/L$ that we find here with respect to those by \citet{DeVitaBZ2016} is in line with the difference among the global values of $M/L$ listed in Table~\ref{Tab_Cfr_Literature}.

\citet{Arca-Sedda2016} proposed an analysis of numerical simulations showing that the excess of mass in the centre of a cluster could be due to the presence of a subsystem of heavy remnants orbitally segregated, and not to an IMBH. In particular, for the globular cluster $\omega$ Cen they estimate a mass of $(1.45 \pm 0.03) \times 10^3$ M$_{\sun}$  for the central component of massive remnants, assuming a total mass of $2.5 \times 10^6$ M$_{\sun}$ for the cluster. Our models predict a larger value for the mass of the dark component, but in our case this is not completely segregated in the centre, extending up to about the half-mass radius. We note that with our choice of $\delta$ for the models, motivated by $N$-body models \citep{Peuten2017}, the distribution of the massive objects is not a choice, but is constrained by the definition of the distribution function, whereas with other methods there is no constraint in this regard. It is however interesting to note that for the dark component in our best-fit models (see Table~\ref{Tab2}) the mass contained within the central 0.2 pc equals $\sim 1.4 \times 10^3$ M$_{\sun}$, in agreement with the estimate by \citet{Arca-Sedda2016}.

\begin{figure}
\centering
\includegraphics[width=0.48\textwidth]{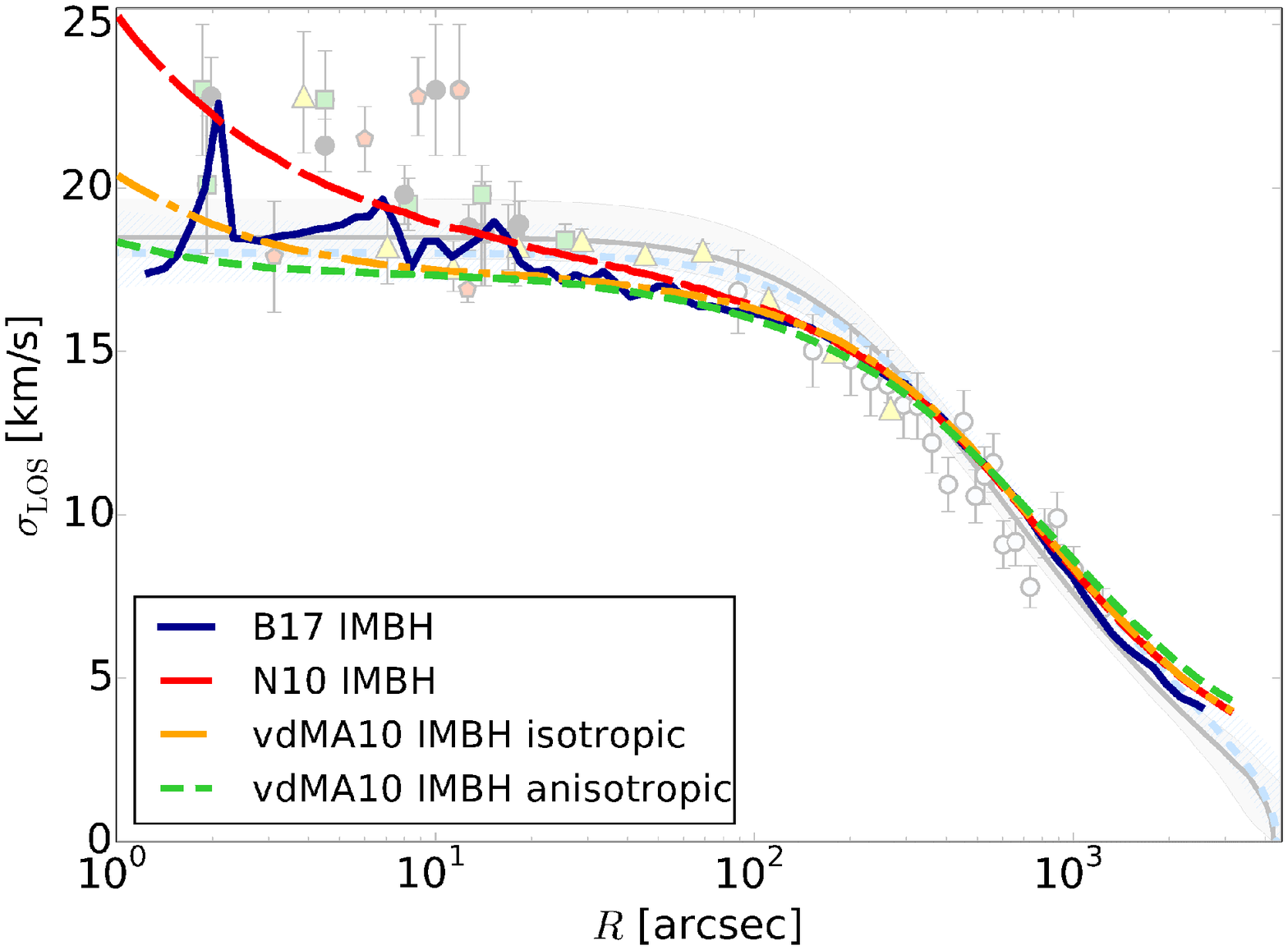}
\caption{Best-fit models including a central IMBH, representing the projected velocity dispersion profile of $\omega$ Cen. The blue solid line represents the best-fit model by \citet{Baumgardt2017}, which includes a central IMBH of mass $4.1 \times 10^4$ M$_{\sun}$. The other models were presented by \citet{vdMA2010}: the red long-dashed line is the best-fit isotropic model including an IMBH of $4 \times 10^4$ M$_{\sun}$ as reported by \citet{Noyola2008}; the orange dot-dashed line is the cusped isotropic model including an IMBH of mass $1.8 \times 10^4$ M$_{\sun}$, and the green dashed line the cusped anisotropic model including an IMBH with mass $8.7 \times 10^3$ M$_{\sun}$. To enable an easier comparison with the models presented here, in the background we reproduced the lines and data points showed in the top right panel of Fig.~\ref{Fig_1S}.}
\label{Fig_cfr_models}
\end{figure}

\subsection{Implications on the presence of an IMBH}

The presence of a central IMBH is accounted for by the analyses carried out by \citet{vdMA2010} and \citet{Baumgardt2017}. \citet{vdMA2010} find an upper limit to the presence of an IMBH in the centre of $\omega$ Cen of $< 1.2 \times 10^4$ M$_{\sun}$, which corresponds to $0.4\%$ of the total mass they estimate for the cluster. The best-fit model proposed by \citet{Baumgardt2017} includes an IMBH amounting to $4.1 \times 10^4$ M$_{\sun}$, i.e. $\sim1.4\%$ of the mass of the cluster. The invisible mass accounted for by these objects is smaller than the one we find for the population of BHs, but much more concentrated. 

Figure~\ref{Fig_cfr_models} shows a comparison between the best-fit models obtained here and some models proposed in the literature including a central IMBH. Even though these models were obtained by fitting to different data, it is interesting to compare them, because they are all describing $\omega$ Cen. The blue solid line in the figure represents the best-fit model by \citet{Baumgardt2017}, which includes a central IMBH of mass $4.1 \times 10^4$ M$_{\sun}$: this model has been obtained by comparing a suite of $N$-body simulations to the observed profiles of the cluster, and it is shown to provide a good fit to the data. In their Fig. 7, \citet{vdMA2010} show several models for $\omega$ Cen, and we reproduce three of these in Fig.~\ref{Fig_cfr_models}, considering only the ones obtained by including the presence of a cusp in the density profile of the cluster: the red long-dashed line is the best-fit isotropic model including an IMBH of $4 \times 10^4$ M$_{\sun}$, as reported by \citet{Noyola2008}; the orange dot-dashed line is the isotropic model including an IMBH of mass $1.8 \times 10^4$ M$_{\sun}$, and the green dashed line the anisotropic model including an IMBH with mass $8.7 \times 10^3$ M$_{\sun}$.

By comparing these models with the ones proposed here, and reproduced in the background of Fig.~\ref{Fig_cfr_models}, we see that a discrepancy is visible in the innermost region of the cluster. The isotropic model by \citet{vdMA2010}, indicated with the dot-dashed orange line in the figure, is above the $1 \sigma$ region of our best-fit two-components models only within $\sim 1 \arcsec$ from the centre, the profile by \citet{Baumgardt2017} is above the $1 \sigma$ region within $\sim 3 \arcsec$ from the centre: it is remarkable that our models, which do not include an IMBH, produce a projected velocity dispersion profile that is indeed very similar to these ones, except for the innermost region, where data are not available. This is an illustration of the partial degeneracy between the signatures produced by an IMBH and by a population of centrally concentrated BHs, and suggests that to firmly assess whether an IMBH is indeed hosted in this system, more data are necessary in the very centre of the cluster.

A larger discrepancy is found when considering the model including the IMBH as suggested by \citet{Noyola2008}, which appears to be above the $1 \sigma$ region within $\sim 8 \arcsec$ from the cluster centre. We point out that, however, this model accounts for a mass of the central IMBH which is basically the same as the model by \citet{Baumgardt2017}, even though the respective profiles look different. Another discrepancy is visible in the intermediate part of the cluster, between $\sim 30$ and $200 \arcsec$, where the models from the literature appear to be below our best-fit models (but still within the $1 \sigma$ region of the isotropic model we presented). This is probably due to the different set of data we considered in the fitting procedure.


\section{Conclusions}
\label{Sect_Concl}

One of the expected signatures of the presence of an IMBH in the centre of globular clusters is a central cusp in the velocity dispersion profiles \citep{Noyola2010,AvdM2010}, and a general increase of the projected velocity dispersion over a region of about an order of magnitude larger than the radius of influence of the IMBH \citep{Baumgardt2004}. A similar feature, however, is also obtained in radially anisotropic stellar systems or in systems having a population of dark heavy remnants. We explored the effect of the first of these alternatives in a previous work \citep{Zocchi2017}, and here we investigated the second.

Stellar-mass BH candidates have been detected in globular clusters \citep{Strader2012b,Chomiuketal2013,MillerJonesetal2015}. From a theoretical point of view, they are also expected to be found in these systems \citep{BreenHeggie2013,Sippel2013,Morscher2015}. For many aspects, the effect of a population of BHs is the same as the one of an IMBH (see for example \citealt{LuetzBK2013} and the dispute between \citealt{Newell1976} and \citealt{IllingworthKing1977}). In particular, their presence quenches mass segregation among the visible stars \citep{Peuten2016,Gill2008}, and clusters containing these objects are expected to have a large core \citep{Heggie2007,Peuten2017}. Moreover, their presence also affects the central velocity dispersion, causing it to be larger than in systems where they do not play a role. Therefore, it is necessary to understand the role of a population of BHs in shaping the dynamics of visible stars, before being able to determine whether or not an IMBH is present in the centre of these systems.

Here we explore this issue by using $\omega$ Cen as a test case. We describe this cluster by means of \limepy\ dynamical models \citep{GielesZocchi} with two mass components: a low-mass component representing the stars and low-mass stellar remnants (white dwarfs and neutron stars) and a high-mass component representing a population of BHs segregated towards the centre. BHs are assumed to be on average more massive than the visible stars, but their total mass is smaller than that of stars. We compare observational profiles to those predicted by the models for the visible component: the dark component has no direct role in the comparison to observations, but its presence influences the dynamics and modifies the shape of these kinematic and structural profiles.  

In order to explore this, we carried out a two-steps fitting procedure by considering models with a given mass function, with the values of the mean and total mass of stars in each component chosen to fully explore the parameter space. The first step consists in a fit to the surface brightness of $\omega$ Cen to determine the values of the model parameters and of the physical scales. With a second step, by using the line-of-sight velocity dispersion, we find the best-fit value of the mass-to-light ratio. This procedure allows us to compare different dynamical models that reproduce the surface brightness profile of $\omega$ Cen in a remarkably similar way. We found that models including a larger component made of more massive BHs are better suited to reproduce the kinematics of this cluster. 

We also carried out an additional fit, without fixing a priori the mass function of the system, and by considering all the observational profiles at once. We did this twice, once by considering isotropic models, and once by also determining the amount of radial anisotropy in the system. The anisotropic models with two mass components perform better with respect to the isotropic ones, as shown by the values of the $\widetilde{\chi}^2$ in Table~\ref{Tab2}, and by the comparison to the projected anisotropy profile measured by proper motions (see Fig.~\ref{Fig_beta_cfr}).

The innermost part of the line-of-sight velocity dispersion profile \citep{Noyola2010} has been used to claim the presence of an IMBH in the centre of this cluster. The best-fit models we obtained with the one-step fitting procedure (and also some of those obtained with the two-step procedure, when considering a large population of heavy BHs) are partially able to reproduce this behaviour. Particularly interesting is the fact that anisotropic single-mass models and isotropic multimass models predict a very similar value for the central velocity dispersion; when combining anisotropy and multiple mass components, again, it is possible to obtain the same value for $\sigma_{0,1}$, but with less extreme mass functions and milder anisotropy. The discrepancy between these best-fit models and the central cusp prevents us from excluding the presence of a central IMBH, but significantly reduces the expected mass, with respect to the one predicted when using single-mass isotropic models \citep[see also][]{Zocchi2017}. We are currently developing a new version of the models proposed here, including the presence of a central IMBH, in order to provide a global description of the dynamics of the stars in GCs, and to estimate the mass of IMBHs that could reside in their centre.

Upcoming measurements of proper motions by the Gaia mission and measurements of line-of-sight velocities by ground-based facilities will soon enable us to study the kinematics of many Galactic globular clusters in great detail and up to their outermost regions, which are currently not explored because data are not available (for $\omega$ Cen, for example, Gaia data will provide proper motions of stars located beyond $\sim 100$ arcsec from the centre, thus complementing HST data). This will enable a more detailed exploration of the kinematics of stars in these systems, and will provide us with information on their invisible components. In particular, the expected errors on proper motions (in Gaia Data Release 2, proper motions for individual stars with $V \leq 18$ mag in $\omega$ Cen will have errors $\leq 3$ km$\,$s$^{-1}$) guarantee that Gaia data will enable us to quantify anisotropy in the outer parts of Galactic globular clusters, lifting part of the degeneracy between anisotropy and heavy remnants.

Both radial anisotropy and the presence of a population of black holes have an important effect on the dynamics of clusters and on their observational properties. In order to provide an accurate estimate for the mass of a central IMBH, it is therefore crucial to take these ingredients into account.

\section*{Acknowledgements}
We thank Antonio Sollima and Sebastian Kamann for comments on an earlier version of the manuscript, and Anna Lisa Varri, Elena Pancino, Anna Sippel, and Eduardo Balbinot for interesting discussions. We thank the anonymous referee for a constructive report that allowed us to improve the paper. AZ acknowledges financial support from the Royal Society (Newton International Fellowship follow on funding). MG acknowledges financial support from the Royal Society (University Research Fellowship) and the European Research Council (ERC StG-335936, CLUSTERS). VHB acknowledges support from the NRC-Canada Plaskett Fellowship and from the Radboud Excellence Initiative.


\bibliographystyle{mn2e}
\bibliography{biblio.bib}

\begin{thebibliography}{}

\bibitem[\protect\citeauthoryear{{Alessandrini}, {Lanzoni}, {Ferraro},
  {Miocchi} \& {Vesperini}}{{Alessandrini} et~al.}{2016}]{Alessandrini2016}
{Alessandrini} E.,  {Lanzoni} B.,  {Ferraro} F.~R.,  {Miocchi} P.,
  {Vesperini} E.,  2016, \apj, 833, 252

\bibitem[\protect\citeauthoryear{{Anderson}}{{Anderson}}{2002}]{Anderson2002}
{Anderson} J.,  2002, in {van Leeuwen} F.,  {Hughes} J.~D.,   {Piotto} G.,
  eds, Omega Centauri, A Unique Window into Astrophysics Vol.~265 of
  Astronomical Society of the Pacific Conference Series, {Main-Sequence
  Observations with HST}.
p.~87

\bibitem[\protect\citeauthoryear{{Anderson} \& {van der Marel}}{{Anderson} \&
  {van der Marel}}{2010}]{AvdM2010}
{Anderson} J.,  {van der Marel} R.~P.,  2010, \apj, 710, 1032

\bibitem[\protect\citeauthoryear{{Arca-Sedda}}{{Arca-Sedda}}{2016}]{Arca-Sedda2016}
{Arca-Sedda} M.,  2016, \mnras, 455, 35

\bibitem[\protect\citeauthoryear{{Baumgardt}}{{Baumgardt}}{2017}]{Baumgardt2017}
{Baumgardt} H.,  2017, \mnras, 464, 2174

\bibitem[\protect\citeauthoryear{{Baumgardt}, {Portegies Zwart}, {McMillan},
  {Makino} \& {Ebisuzaki}}{{Baumgardt} et~al.}{2004}]{Baumgardt2004}
{Baumgardt} H.,  {Portegies Zwart} S.~F.,  {McMillan} S.~L.~W.,  {Makino} J.,
   {Ebisuzaki} T.,  2004, in {Lamers} H.~J.~G.~L.~M.,  {Smith} L.~J.,   {Nota}
  A.,  eds, The Formation and Evolution of Massive Young Star Clusters Vol.~322
  of Astronomical Society of the Pacific Conference Series, {Dynamics of
  Intermediate Mass Black Holes in Star Clusters}.
p.~459

\bibitem[\protect\citeauthoryear{{Bellini}, {Libralato}, {Bedin}, {Milone},
  {van der Marel}, {Anderson}, {Apai}, {Burgasser}, {Marino} \&
  {Rees}}{{Bellini} et~al.}{2018}]{Bellini2018}
{Bellini} A.,  {Libralato} M.,  {Bedin} L.~R.,  {Milone} A.~P.,  {van der
  Marel} R.~P.,  {Anderson} J.,  {Apai} D.,  {Burgasser} A.~J.,  {Marino}
  A.~F.,    {Rees} J.~M.,  2018, \apj, 853, 86

\bibitem[\protect\citeauthoryear{{Bianchini}, {van de Ven}, {Norris},
  {Schinnerer} \& {Varri}}{{Bianchini} et~al.}{2016}]{Bianchini2016}
{Bianchini} P.,  {van de Ven} G.,  {Norris} M.~A.,  {Schinnerer} E.,    {Varri}
  A.~L.,  2016, \mnras, 458, 3644

\bibitem[\protect\citeauthoryear{{Bianchini}, {Varri}, {Bertin} \&
  {Zocchi}}{{Bianchini} et~al.}{2013}]{BVBZ2013}
{Bianchini} P.,  {Varri} A.~L.,  {Bertin} G.,    {Zocchi} A.,  2013, \apj, 772,
  67

\bibitem[\protect\citeauthoryear{{Breen} \& {Heggie}}{{Breen} \&
  {Heggie}}{2013}]{BreenHeggie2013}
{Breen} P.~G.,  {Heggie} D.~C.,  2013, \mnras, 432, 2779

\bibitem[\protect\citeauthoryear{{Chomiuk}, {Strader}, {Maccarone},
  {Miller-Jones}, {Heinke}, {Noyola}, {Seth} \& {Ransom}}{{Chomiuk}
  et~al.}{2013}]{Chomiuketal2013}
{Chomiuk} L.,  {Strader} J.,  {Maccarone} T.~J.,  {Miller-Jones} J.~C.~A.,
  {Heinke} C.,  {Noyola} E.,  {Seth} A.~C.,    {Ransom} S.,  2013, \apj, 777,
  69

\bibitem[\protect\citeauthoryear{{Da Costa} \& {Freeman}}{{Da Costa} \&
  {Freeman}}{1976}]{DaCostaFreeman1976}
{Da Costa} G.~S.,  {Freeman} K.~C.,  1976, \apj, 206, 128

\bibitem[\protect\citeauthoryear{{Dalessandro}, {Ferraro}, {Massari},
  {Lanzoni}, {Miocchi} \& {Beccari}}{{Dalessandro} et~al.}{2015}]{D15}
{Dalessandro} E.,  {Ferraro} F.~R.,  {Massari} D.,  {Lanzoni} B.,  {Miocchi}
  P.,    {Beccari} G.,  2015, \apj, 810, 40

\bibitem[\protect\citeauthoryear{{de Vita}, {Bertin} \& {Zocchi}}{{de Vita}
  et~al.}{2016}]{DeVitaBZ2016}
{de Vita} R.,  {Bertin} G.,    {Zocchi} A.,  2016, \aap, 590, A16

\bibitem[\protect\citeauthoryear{{Djorgovski} \& {King}}{{Djorgovski} \&
  {King}}{1986}]{DjorgovskiKing1986}
{Djorgovski} S.,  {King} I.~R.,  1986, \apjl, 305, L61

\bibitem[\protect\citeauthoryear{{Ebisuzaki}, {Makino}, {Tsuru}, {Funato},
  {Portegies Zwart}, {Hut}, {McMillan}, {Matsushita}, {Matsumoto} \&
  {Kawabe}}{{Ebisuzaki} et~al.}{2001}]{Ebisuzaki2001}
{Ebisuzaki} T.,  {Makino} J.,  {Tsuru} T.~G.,  {Funato} Y.,  {Portegies Zwart}
  S.,  {Hut} P.,  {McMillan} S.,  {Matsushita} S.,  {Matsumoto} H.,    {Kawabe}
  R.,  2001, \apjl, 562, L19

\bibitem[\protect\citeauthoryear{{Feldmeier}, {L{\"u}tzgendorf}, {Neumayer},
  {Kissler-Patig}, {Gebhardt}, {Baumgardt}, {Noyola}, {de Zeeuw} \&
  {Jalali}}{{Feldmeier} et~al.}{2013}]{Feldmeier2013}
{Feldmeier} A.,  {L{\"u}tzgendorf} N.,  {Neumayer} N.,  {Kissler-Patig} M.,
  {Gebhardt} K.,  {Baumgardt} H.,  {Noyola} E.,  {de Zeeuw} P.~T.,    {Jalali}
  B.,  2013, \aap, 554, A63

\bibitem[\protect\citeauthoryear{{Ferraro}, {Sollima}, {Rood}, {Origlia},
  {Pancino} \& {Bellazzini}}{{Ferraro} et~al.}{2006}]{Ferraro2006}
{Ferraro} F.~R.,  {Sollima} A.,  {Rood} R.~T.,  {Origlia} L.,  {Pancino} E.,
  {Bellazzini} M.,  2006, \apj, 638, 433

\bibitem[\protect\citeauthoryear{{Foreman-Mackey}, {Hogg}, {Lang} \&
  {Goodman}}{{Foreman-Mackey} et~al.}{2013}]{emcee_paper}
{Foreman-Mackey} D.,  {Hogg} D.~W.,  {Lang} D.,    {Goodman} J.,  2013, PASP,
  125, 306

\bibitem[\protect\citeauthoryear{{Fridman}, {Polyachenko}, {Aries} \&
  {Poliakoff}}{{Fridman} et~al.}{1984}]{FridmanPolyachenko1984}
{Fridman} A.~M.,  {Polyachenko} V.~L.,  {Aries} A.~B.,    {Poliakoff} I.~N.,
  1984, {Physics of gravitating systems. I. Equilibrium and stability.}

\bibitem[\protect\citeauthoryear{{Gieles}, {Charbonnel}, {Krause},
  {Henault-Brunet}, {Agertz}, {Lamers}, {Bastian}, {Gualandris}, {Zocchi} \&
  {Petts}}{{Gieles} et~al.}{2018}]{Gieles2018SMS}
{Gieles} M.,  {Charbonnel} C.,  {Krause} M.,  {Henault-Brunet} V.,  {Agertz}
  O.,  {Lamers} H.,  {Bastian} N.,  {Gualandris} A.,  {Zocchi} A.,    {Petts}
  J.,  2018, ArXiv e-prints

\bibitem[\protect\citeauthoryear{{Gieles} \& {Zocchi}}{{Gieles} \&
  {Zocchi}}{2015}]{GielesZocchi}
{Gieles} M.,  {Zocchi} A.,  2015, \mnras, 454, 576

\bibitem[\protect\citeauthoryear{{Giersz} \& {Heggie}}{{Giersz} \&
  {Heggie}}{2003}]{GierszHeggie2003}
{Giersz} M.,  {Heggie} D.~C.,  2003, \mnras, 339, 486

\bibitem[\protect\citeauthoryear{{Gill}, {Trenti}, {Miller}, {van der Marel},
  {Hamilton} \& {Stiavelli}}{{Gill} et~al.}{2008}]{Gill2008}
{Gill} M.,  {Trenti} M.,  {Miller} M.~C.,  {van der Marel} R.,  {Hamilton} D.,
    {Stiavelli} M.,  2008, \apj, 686, 303

\bibitem[\protect\citeauthoryear{{Gomez-Leyton} \& {Velazquez}}{{Gomez-Leyton}
  \& {Velazquez}}{2014}]{G-L_V2014}
{Gomez-Leyton} Y.~J.,  {Velazquez} L.,  2014, Journal of Statistical Mechanics:
  Theory and Experiment, 4, 6

\bibitem[\protect\citeauthoryear{{Gunn} \& {Griffin}}{{Gunn} \&
  {Griffin}}{1979}]{GunnGriffin1979}
{Gunn} J.~E.,  {Griffin} R.~F.,  1979, \aj, 84, 752

\bibitem[\protect\citeauthoryear{{Haggard}, {Cool}, {Heinke}, {van der Marel},
  {Cohn}, {Lugger} \& {Anderson}}{{Haggard} et~al.}{2013}]{Haggard2013}
{Haggard} D.,  {Cool} A.~M.,  {Heinke} C.~O.,  {van der Marel} R.,  {Cohn}
  H.~N.,  {Lugger} P.~M.,    {Anderson} J.,  2013, \apjl, 773, L31

\bibitem[\protect\citeauthoryear{{Heggie}, {Hut}, {Mineshige}, {Makino} \&
  {Baumgardt}}{{Heggie} et~al.}{2007}]{Heggie2007}
{Heggie} D.~C.,  {Hut} P.,  {Mineshige} S.,  {Makino} J.,    {Baumgardt} H.,
  2007, \pasj, 59, L11

\bibitem[\protect\citeauthoryear{{Hurley}, {Pols} \& {Tout}}{{Hurley}
  et~al.}{2000}]{Hurley2000}
{Hurley} J.~R.,  {Pols} O.~R.,    {Tout} C.~A.,  2000, \mnras, 315, 543

\bibitem[\protect\citeauthoryear{{Illingworth} \& {King}}{{Illingworth} \&
  {King}}{1977}]{IllingworthKing1977}
{Illingworth} G.,  {King} I.~R.,  1977, \apjl, 218, L109

\bibitem[\protect\citeauthoryear{{Kamann}, {Husser}, {Dreizler}, {Emsellem},
  {Weilbacher}, {Martens}, {Bacon}, {den Brok}, {Giesers}, {Krajnovi{\'c}},
  {Roth}, {Wendt} \& {Wisotzki}}{{Kamann} et~al.}{2018}]{Kamann2018}
{Kamann} S.,  {Husser} T.-O.,  {Dreizler} S.,  {Emsellem} E.,  {Weilbacher}
  P.~M.,  {Martens} S.,  {Bacon} R.,  {den Brok} M.,  {Giesers} B.,
  {Krajnovi{\'c}} D.,  {Roth} M.~M.,  {Wendt} M.,    {Wisotzki} L.,  2018,
  \mnras, 473, 5591

\bibitem[\protect\citeauthoryear{{King}}{{King}}{1966}]{King1966}
{King} I.~R.,  1966, \aj, 71, 64

\bibitem[\protect\citeauthoryear{{Kroupa}, {Aarseth} \& {Hurley}}{{Kroupa}
  et~al.}{2001}]{Kroupa2001}
{Kroupa} P.,  {Aarseth} S.,    {Hurley} J.,  2001, \mnras, 321, 699

\bibitem[\protect\citeauthoryear{{Lu} \& {Kong}}{{Lu} \&
  {Kong}}{2011}]{LuKong2011}
{Lu} T.-N.,  {Kong} A.~K.~H.,  2011, \apjl, 729, L25

\bibitem[\protect\citeauthoryear{{L{\"u}tzgendorf}, {Baumgardt} \&
  {Kruijssen}}{{L{\"u}tzgendorf} et~al.}{2013}]{LuetzBK2013}
{L{\"u}tzgendorf} N.,  {Baumgardt} H.,    {Kruijssen} J.~M.~D.,  2013, \aap,
  558, A117

\bibitem[\protect\citeauthoryear{{L{\"u}tzgendorf}, {Kissler-Patig},
  {Gebhardt}, {Baumgardt}, {Noyola}, {de Zeeuw}, {Neumayer}, {Jalali} \&
  {Feldmeier}}{{L{\"u}tzgendorf} et~al.}{2013}]{Luetz2013}
{L{\"u}tzgendorf} N.,  {Kissler-Patig} M.,  {Gebhardt} K.,  {Baumgardt} H.,
  {Noyola} E.,  {de Zeeuw} P.~T.,  {Neumayer} N.,  {Jalali} B.,    {Feldmeier}
  A.,  2013, \aap, 552, A49

\bibitem[\protect\citeauthoryear{{L{\"u}tzgendorf}, {Kissler-Patig}, {Noyola},
  {Jalali}, {de Zeeuw}, {Gebhardt} \& {Baumgardt}}{{L{\"u}tzgendorf}
  et~al.}{2011}]{Nora2011}
{L{\"u}tzgendorf} N.,  {Kissler-Patig} M.,  {Noyola} E.,  {Jalali} B.,  {de
  Zeeuw} P.~T.,  {Gebhardt} K.,    {Baumgardt} H.,  2011, \aap, 533, A36

\bibitem[\protect\citeauthoryear{{Maccarone}, {Kundu}, {Zepf} \&
  {Rhode}}{{Maccarone} et~al.}{2007}]{Maccarone2007}
{Maccarone} T.~J.,  {Kundu} A.,  {Zepf} S.~E.,    {Rhode} K.~L.,  2007, \nat,
  445, 183

\bibitem[\protect\citeauthoryear{{Mackey}, {Wilkinson}, {Davies} \&
  {Gilmore}}{{Mackey} et~al.}{2008}]{Mackeyetal2008}
{Mackey} A.~D.,  {Wilkinson} M.~I.,  {Davies} M.~B.,    {Gilmore} G.~F.,  2008,
  \mnras, 386, 65

\bibitem[\protect\citeauthoryear{{Mandel}}{{Mandel}}{2016}]{Mandel2016}
{Mandel} I.,  2016, \mnras, 456, 578

\bibitem[\protect\citeauthoryear{{Merritt}, {Piatek}, {Portegies Zwart} \&
  {Hemsendorf}}{{Merritt} et~al.}{2004}]{Merrittetal2004}
{Merritt} D.,  {Piatek} S.,  {Portegies Zwart} S.,    {Hemsendorf} M.,  2004,
  \apjl, 608, L25

\bibitem[\protect\citeauthoryear{{Meylan}}{{Meylan}}{1987}]{Meylan1987}
{Meylan} G.,  1987, \aap, 184, 144

\bibitem[\protect\citeauthoryear{{Meylan}, {Mayor}, {Duquennoy} \&
  {Dubath}}{{Meylan} et~al.}{1995}]{Meylan1995}
{Meylan} G.,  {Mayor} M.,  {Duquennoy} A.,    {Dubath} P.,  1995, \aap, 303,
  761

\bibitem[\protect\citeauthoryear{{Michie}}{{Michie}}{1963}]{Michie1963}
{Michie} R.~W.,  1963, \mnras, 125, 127

\bibitem[\protect\citeauthoryear{{Miller-Jones}, {Strader}, {Heinke},
  {Maccarone}, {van den Berg}, {Knigge}, {Chomiuk}, {Noyola}, {Russell}, {Seth}
  \& {Sivakoff}}{{Miller-Jones} et~al.}{2015}]{MillerJonesetal2015}
{Miller-Jones} J.~C.~A.,  {Strader} J.,  {Heinke} C.~O.,  {Maccarone} T.~J.,
  {van den Berg} M.,  {Knigge} C.,  {Chomiuk} L.,  {Noyola} E.,  {Russell}
  T.~D.,  {Seth} A.~C.,    {Sivakoff} G.~R.,  2015, \mnras, 453, 3918

\bibitem[\protect\citeauthoryear{{Morscher}, {Pattabiraman}, {Rodriguez},
  {Rasio} \& {Umbreit}}{{Morscher} et~al.}{2015}]{Morscher2015}
{Morscher} M.,  {Pattabiraman} B.,  {Rodriguez} C.,  {Rasio} F.~A.,
  {Umbreit} S.,  2015, \apj, 800, 9

\bibitem[\protect\citeauthoryear{{Morscher}, {Umbreit}, {Farr} \&
  {Rasio}}{{Morscher} et~al.}{2013}]{Morscher2013}
{Morscher} M.,  {Umbreit} S.,  {Farr} W.~M.,    {Rasio} F.~A.,  2013, \apjl,
  763, L15

\bibitem[\protect\citeauthoryear{{Newell}, {Da Costa} \& {Norris}}{{Newell}
  et~al.}{1976}]{Newell1976}
{Newell} B.,  {Da Costa} G.~S.,    {Norris} J.,  1976, \apjl, 208, L55

\bibitem[\protect\citeauthoryear{{Noyola}, {Gebhardt} \& {Bergmann}}{{Noyola}
  et~al.}{2008}]{Noyola2008}
{Noyola} E.,  {Gebhardt} K.,    {Bergmann} M.,  2008, \apj, 676, 1008

\bibitem[\protect\citeauthoryear{{Noyola}, {Gebhardt}, {Kissler-Patig},
  {L{\"u}tzgendorf}, {Jalali}, {de Zeeuw} \& {Baumgardt}}{{Noyola}
  et~al.}{2010}]{Noyola2010}
{Noyola} E.,  {Gebhardt} K.,  {Kissler-Patig} M.,  {L{\"u}tzgendorf} N.,
  {Jalali} B.,  {de Zeeuw} P.~T.,    {Baumgardt} H.,  2010, \apjl, 719, L60

\bibitem[\protect\citeauthoryear{{Pancino}, {Galfo}, {Ferraro} \&
  {Bellazzini}}{{Pancino} et~al.}{2007}]{Pancino2007}
{Pancino} E.,  {Galfo} A.,  {Ferraro} F.~R.,    {Bellazzini} M.,  2007, \apjl,
  661, L155

\bibitem[\protect\citeauthoryear{{Peuten}, {Zocchi}, {Gieles}, {Gualandris} \&
  {H{\'e}nault-Brunet}}{{Peuten} et~al.}{2016}]{Peuten2016}
{Peuten} M.,  {Zocchi} A.,  {Gieles} M.,  {Gualandris} A.,
  {H{\'e}nault-Brunet} V.,  2016, \mnras, 462, 2333

\bibitem[\protect\citeauthoryear{{Peuten}, {Zocchi}, {Gieles} \&
  {H{\'e}nault-Brunet}}{{Peuten} et~al.}{2017}]{Peuten2017}
{Peuten} M.,  {Zocchi} A.,  {Gieles} M.,    {H{\'e}nault-Brunet} V.,  2017,
  \mnras\ submitted, arXiv:1702.01712

\bibitem[\protect\citeauthoryear{{Polyachenko} \& {Shukhman}}{{Polyachenko} \&
  {Shukhman}}{1981}]{PolyachenkoShukhman1981}
{Polyachenko} V.~L.,  {Shukhman} I.~G.,  1981, \sovast, 25, 533

\bibitem[\protect\citeauthoryear{{Portegies Zwart}, {Baumgardt}, {Hut},
  {Makino} \& {McMillan}}{{Portegies Zwart} et~al.}{2004}]{PortegiesZwart2004}
{Portegies Zwart} S.~F.,  {Baumgardt} H.,  {Hut} P.,  {Makino} J.,
  {McMillan} S.~L.~W.,  2004, \nat, 428, 724

\bibitem[\protect\citeauthoryear{{Reijns}, {Seitzer}, {Arnold}, {Freeman},
  {Ingerson}, {van den Bosch}, {van de Ven} \& {de Zeeuw}}{{Reijns}
  et~al.}{2006}]{Reijns2006}
{Reijns} R.~A.,  {Seitzer} P.,  {Arnold} R.,  {Freeman} K.~C.,  {Ingerson} T.,
  {van den Bosch} R.~C.~E.,  {van de Ven} G.,    {de Zeeuw} P.~T.,  2006, \aap,
  445, 503

\bibitem[\protect\citeauthoryear{{Repetto}, {Davies} \& {Sigurdsson}}{{Repetto}
  et~al.}{2012}]{Repettoetal2012}
{Repetto} S.,  {Davies} M.~B.,    {Sigurdsson} S.,  2012, \mnras, 425, 2799

\bibitem[\protect\citeauthoryear{{Repetto}, {Igoshev} \& {Nelemans}}{{Repetto}
  et~al.}{2017}]{Repetto2017}
{Repetto} S.,  {Igoshev} A.~P.,    {Nelemans} G.,  2017, \mnras, 467, 298

\bibitem[\protect\citeauthoryear{{Repetto} \& {Nelemans}}{{Repetto} \&
  {Nelemans}}{2015}]{Repetto2015}
{Repetto} S.,  {Nelemans} G.,  2015, \mnras, 453, 3341

\bibitem[\protect\citeauthoryear{{Shanahan} \& {Gieles}}{{Shanahan} \&
  {Gieles}}{2015}]{ShanahanGieles2015}
{Shanahan} R.~L.,  {Gieles} M.,  2015, \mnras, 448, L94

\bibitem[\protect\citeauthoryear{{Sippel} \& {Hurley}}{{Sippel} \&
  {Hurley}}{2013}]{Sippel2013}
{Sippel} A.~C.,  {Hurley} J.~R.,  2013, \mnras, 430, L30

\bibitem[\protect\citeauthoryear{{Spera}, {Mapelli} \& {Bressan}}{{Spera}
  et~al.}{2015}]{Spera2015}
{Spera} M.,  {Mapelli} M.,    {Bressan} A.,  2015, \mnras, 451, 4086

\bibitem[\protect\citeauthoryear{{Spera}, {Mapelli} \& {Jeffries}}{{Spera}
  et~al.}{2016}]{Spera2016}
{Spera} M.,  {Mapelli} M.,    {Jeffries} R.~D.,  2016, \mnras, 460, 317

\bibitem[\protect\citeauthoryear{{Spitzer}}{{Spitzer}}{1987}]{Spitzer1987}
{Spitzer} L.,  1987, {Dynamical evolution of globular clusters}.
Princeton University Press, Princeton, NJ

\bibitem[\protect\citeauthoryear{{Spitzer}}{{Spitzer}}{1969}]{Spitzer1969}
{Spitzer} L.~J.,  1969, \apjl, 158, L139

\bibitem[\protect\citeauthoryear{{Strader}, {Chomiuk}, {Maccarone},
  {Miller-Jones} \& {Seth}}{{Strader} et~al.}{2012}]{Strader2012b}
{Strader} J.,  {Chomiuk} L.,  {Maccarone} T.~J.,  {Miller-Jones} J.~C.~A.,
  {Seth} A.~C.,  2012, \nat, 490, 71

\bibitem[\protect\citeauthoryear{{Strader}, {Chomiuk}, {Maccarone},
  {Miller-Jones}, {Seth}, {Heinke} \& {Sivakoff}}{{Strader}
  et~al.}{2012}]{Strader2012}
{Strader} J.,  {Chomiuk} L.,  {Maccarone} T.~J.,  {Miller-Jones} J.~C.~A.,
  {Seth} A.~C.,  {Heinke} C.~O.,    {Sivakoff} G.~R.,  2012, \apjl, 750, L27

\bibitem[\protect\citeauthoryear{{Trager}, {King} \& {Djorgovski}}{{Trager}
  et~al.}{1995}]{TKD1995}
{Trager} S.~C.,  {King} I.~R.,    {Djorgovski} S.,  1995, \aj, 109, 218

\bibitem[\protect\citeauthoryear{{Trenti} \& {van der Marel}}{{Trenti} \& {van
  der Marel}}{2013}]{TrentivdM2013}
{Trenti} M.,  {van der Marel} R.,  2013, \mnras, 435, 3272

\bibitem[\protect\citeauthoryear{{van de Ven}, {van den Bosch}, {Verolme} \&
  {de Zeeuw}}{{van de Ven} et~al.}{2006}]{vandeVen2006}
{van de Ven} G.,  {van den Bosch} R.~C.~E.,  {Verolme} E.~K.,    {de Zeeuw}
  P.~T.,  2006, \aap, 445, 513

\bibitem[\protect\citeauthoryear{{van der Marel} \& {Anderson}}{{van der Marel}
  \& {Anderson}}{2010}]{vdMA2010}
{van der Marel} R.~P.,  {Anderson} J.,  2010, \apj, 710, 1063

\bibitem[\protect\citeauthoryear{{van Leeuwen}, {Le Poole}, {Reijns}, {Freeman}
  \& {de Zeeuw}}{{van Leeuwen} et~al.}{2000}]{vanLeeuwen2000}
{van Leeuwen} F.,  {Le Poole} R.~S.,  {Reijns} R.~A.,  {Freeman} K.~C.,    {de
  Zeeuw} P.~T.,  2000, \aap, 360, 472

\bibitem[\protect\citeauthoryear{{Varri} \& {Bertin}}{{Varri} \&
  {Bertin}}{2012}]{VarriBertin2012}
{Varri} A.~L.,  {Bertin} G.,  2012, \aap, 540, A94

\bibitem[\protect\citeauthoryear{{Vesperini} \& {Trenti}}{{Vesperini} \&
  {Trenti}}{2010}]{Vesperini2010}
{Vesperini} E.,  {Trenti} M.,  2010, \apjl, 720, L179

\bibitem[\protect\citeauthoryear{{Vishniac}}{{Vishniac}}{1978}]{Vishniac1978}
{Vishniac} E.~T.,  1978, \apj, 223, 986

\bibitem[\protect\citeauthoryear{{Watkins}, {van de Ven}, {den Brok} \& {van
  den Bosch}}{{Watkins} et~al.}{2013}]{Watkins2013}
{Watkins} L.~L.,  {van de Ven} G.,  {den Brok} M.,    {van den Bosch} R.~C.~E.,
   2013, \mnras, 436, 2598

\bibitem[\protect\citeauthoryear{{Watkins}, {van der Marel}, {Bellini} \&
  {Anderson}}{{Watkins} et~al.}{2015}]{Watkins2015}
{Watkins} L.~L.,  {van der Marel} R.~P.,  {Bellini} A.,    {Anderson} J.,
  2015, \apj, 812, 149

\bibitem[\protect\citeauthoryear{{Watters}, {Joshi} \& {Rasio}}{{Watters}
  et~al.}{2000}]{Watters2000}
{Watters} W.~A.,  {Joshi} K.~J.,    {Rasio} F.~A.,  2000, \apj, 539, 331

\bibitem[\protect\citeauthoryear{{Wilson}}{{Wilson}}{1975}]{Wilson1975}
{Wilson} C.~P.,  1975, \aj, 80, 175

\bibitem[\protect\citeauthoryear{{Woolley}}{{Woolley}}{1954}]{Woolley1954}
{Woolley} R.~V.~D.~R.,  1954, \mnras, 114, 191

\bibitem[\protect\citeauthoryear{{Zocchi}, {Gieles} \&
  {H{\'e}nault-Brunet}}{{Zocchi} et~al.}{2017}]{Zocchi2017}
{Zocchi} A.,  {Gieles} M.,    {H{\'e}nault-Brunet} V.,  2017, \mnras\ accepted,
  arXiv:1702.00725

\end{thebibliography}


\label{lastpage}
\end{document}